# Active emulsions in living cell membranes driven by contractile stresses and transbilayer coupling

Suvrajit Saha[1,2,*], Amit Das[1,3,*], Chandrima Patra[1], Anupama Ambika Anilkumar[1,4], Parijat Sil[1,5], Satyajit Mayor[1,†], and Madan Rao[1,†]

[1]National Centre for Biological Sciences (TIFR), Bellary Road, Bangalore 560065, India
[2]Cardiovascular Research Institute, University of California San Francisco (UCSF), San Francisco, California, USA
[3]Northeastern University, Boston, Massachusetts, USA
[4]Present address: Deputy Scientific Attaché, Embassy of France in India
[5]Present address: Department of Molecular Biology, Princeton University, Princeton, NJ, USA
[*]S.S. & A.D. contributed equally to this work. †Corresponding authors: mayor@ncbs.res.in & madan@ncbs.res.in

## Abstract

The spatiotemporal organisation of proteins and lipids on the cell surface has direct functional consequences for signaling, sorting and endocytosis. Earlier studies have shown that multiple types of membrane proteins including transmembrane proteins that have cytoplasmic actin binding capacity and lipid-tethered GPI-anchored proteins (GPI-APs) form nanoscale clusters driven by active contractile flows generated by the actin cortex. To gain insight into the role of lipids in organizing membrane domains in living cells, we study the molecular interactions that promote the actively generated nanoclusters of GPI-APs and transmembrane proteins. This motivates a theoretical description, wherein a combination of active contractile stresses and transbilayer coupling drive the creation of *active emulsions*, mesoscale liquid ordered (lo) domains of the GPI-APs and lipids, at temperatures greater than equilibrium lipid-phase segregation. To test these ideas we use spatial imaging of homo-FRET combined with local membrane order and demonstrate that mesoscopic domains enriched in nanoclusters of GPI-APs are maintained by cortical actin activity and transbilayer interactions, and exhibit significant lipid order, consistent with predictions of the active composite model.

## Introduction

An outstanding issue in modern cell biology is the spatiotemporal organisation of composition. The plasma membrane of a living cell with its diversity of proteins and lipids is an archetypal example (1). Numerous studies have revealed that molecular organisation at the cell surface, ranging from $10 - 10^3$ nm in space and $10^{-3} - 10$ s in time, have direct functional consequences for signaling reactions, molecular sorting and endocytosis (2, 3). Following the fluid-mosaic model (4), several equilibrium models, such as the *lipid-shell* model (5) and the *lipid-raft* model (6), have been proposed to describe the lateral organisation of the cell membrane. More recently, it has been shown that many cell surface proteins and lipids form dynamic nanoclusters driven by the activity of actomyosin at the cell cortex (7–12). This not only identifies a novel molecular organiser of local plasma membrane composition, namely cortical actomyosin (13–15), but also a new mechanism, highlighting the role of non-equilibrium, energy-consuming 'active' processes (16) in driving local compositional heterogeneities (17).

This is the motivation for the *active composite cell surface* model (8, 17–19), a juxtaposition of the membrane bilayer and the actomyosin cortex, where membrane components that couple to the cortical elements are subject to fluctuating actomyosin contractile stresses that drive flows in localized regions (20–22), leading to the formation of dynamic (nano)clusters on the cell surface (7, 8, 22). Predictions from this model have been verified in high resolution fluorescence experiments in living cells (7, 8, 15, 23), and in *in vitro* reconstitutions of the active composite, *viz.* actomyosin layered atop a planar bilayer membrane (19, 24).

The active nanoclustering appears to be a general phenomena (13); it is exhibited by transmembrane proteins that directly bind to actin, such as model proteins that consist of actin binding motifs at their cytoplasmic tails (TMABD; (8)) and by naturally occurring proteins such as E-Cadherin and CD44, that recruit actin-binding modules to bind to cortical actin (10, 11). Further, lipid-tethered membrane proteins, such as outer leaflet GPI-anchored proteins (GPI-AP) (8, 25, 26), glycolipids (12) and inner leaflet Ras-GTPases (27) also exhibit actin-dependent nanoclustering. While direct association with actin provides a mechanism for membrane molecules to couple to the actomyosin cortex, coupling between outer-leaflet GPI-APs or glycolipids and dynamic cortical actin filaments at the inner-leaflet is mediated by transbilayer acyl-chain interactions involving long acyl chain-containing GPI-anchors, inner-leaflet Phosphatidylserine (PS), and cholesterol (28). Sphingolipids (SM) have also been implicated in the maintainence of these nanoclusters (25). A necessary condition for the local stability of the transbilayer interaction at physiological temperatures was found to be the immobilization of PS at the inner leaflet, together with the inter-digitation of long acyl chain lipids across the bilayer and the presence of adequate levels of cholesterol and sphingolipids, predicting a nano-environment with local liquid order (*lo*) (28).

In this paper, we use high resolution fluorescence-based assays to study more closely the role of lipidic interactions, both transbilayer and lateral, in the active organisation of the lipid-tethered GPI-AP and the transmembrane protein, TMABD (Figure 1A schematic). These observations provide the motivation for an active segregation model of a multi-component fluid bilayer membrane comprising outer-leaflet GPI-AP, inner-leaflet PS and *lo* and *ld* preferring lipids in both leaflets. We show that the combination of (i) active stress fluctuations arising from a coupling to cortical actomyosin, (ii) strong transbilayer interaction between PS and GPI / *lo* lipids, and iii) lateral lipidic interactions, drive the mesoscale organisation of GPI-APs at temperatures above the *lo-ld* phase segregation temperature. Such actively segregated mesoscale domains or *active emulsions*, are expected to exhibit anomalous growth dynamics and fluctuations that are unique to the nonequilibrium steady state (29, 30).

To test these theoretical ideas, we use a combination of high-resolution maps of homo-FRET (31) and local membrane order (32), to quantify the mesoscale organisation of GPI-AP and TMABD nanoclusters, and their correlation with the lipid environment. We find mesoscale domains of size $\approx$ $0.1\,\mu\text{m}^2$ enriched in GPI-AP nanoclusters, which in turn co-register with inner-leaflet PS and are associated with higher lipid-order; while mesoscale domains of TMABD nanoclusters are uncorrelated with lipid order. Importantly, disruption of key drivers of activity, *viz*. the dynamic actin-filament nucleator formin, and mediators of GPI-AP binding to actin, *viz*. transbilayer coupling of GPI-AP to PS, lead to a loss of *lo*-mesoscale domains. These results show how the interplay between active stresses (due to actomyosin contractility) and short-range passive forces (here represented by transbilayer coupling and weak lateral lipid-lipid interactions), provide the driving force for the formation of active emulsions at physiological temperatures, qualitatively and quantitatively consistent with predictions from our theoretical model.

## Results

**GPI-anchored proteins and transmembrane proteins form active nanoclusters albeit with different intermolecular interactions.** In earlier studies we and others had shown that GPI-AP and TMABD form nanoclusters (8, 33). Here, using homo-FRET imaging at high resolution to identify nanoclustering of fluorescently labeled proteins ((31, 34), we confirm that GPI-AP and TMABD exhibit actin-dependent nanoscale clusters contingent on the same actin nucleator, formin, rather than Arp2/3 (35)(Fig. 1). Treatment of CHO cells expressing GPI-anchored Folate Receptor (FR-GPI) or FR-tagged-TMABD (FR-TMABD) labeled with N$\alpha$-pteroyl-N$\epsilon$-BodipyTMR- L-lysine [PLB; (8)] with inhibitors of formins [SMIFH2 (36)] but not Arp2/3 [CK666 (37)], disrupted the nanoscale clustering of both FR-GPI (Fig. 1 B, C) and FR-TMABD (Fig. 1 D, E). The cytoplasmic actin-filament binding domain (ABD) allows FR-TMABD to bind directly to actin (8). A single point mutation in the cytosolic ABD (R579A; ABD*) abrogates actin binding and resulted in a complete loss of the formation of actin-driven nanoclusters (8).

The binding of GPI-AP is however indirect, via a transbilayer coupling to phosphatidylserine (PS) at the inner-leaflet and cytoplasmic actin (28). Consequently, perturbations of PS abrogate nanoclustering of GPI-APs (28) but not of FR-TMABD (38). Furthermore, depletion of cholesterol and sphingomyelin (SM) levels alter the nanoscale organisation of GPI-AP(Fig. 1F and Fig. S1A) (23, 25, 39), but have no effect on the nanoclustering of TMABD (Fig. 1G and Fig. S1B), consistent with the idea that both GPI-AP and TMABD engage with the active contractile machinery via distinct molecular interactions.

**Theoretical description of active segregation involving active contractile stresses and transbilayer coupling.** These observations motivate a theoretical description of active segregation of the components of an asymmetric bilayer membrane, that involves combining the nonequilibrium effects of actomyosin contractility with passive lipidic forces, namely the lateral interactions with *lo*-lipid components and transbilayer interactions with lower leaflet PS (Fig. 2A).

Taking a coarse-grained view, we represent the asymmetric bilayer membrane by different components that include the primary players under discussion (Fig. S2A) - (i) GPI-AP in the upper-leaflet, (ii) *lo*-component in the upper-leaflet comprising SM and cholesterol, (iii) lower-leaflet PS together with other long-chain saturated lipids and cholesterol, *ld*-component in the (iv) upper leaflet, and the (v) lower leaflet, comprising short chain or unsaturated lipids and whenever applicable, a sixth component, (vi) transmembrane protein TMABD. The membrane bilayer adjoins a layer of cortical actomyosin, which represents the source of the fluctuating active contractile stresses $\Sigma$ localized over a spatial scale $\xi$ and a lifetime $\tau$ (8, 16, 18), and directly affects the the lower leaflet PS and the transmembrane TMABD, proportional to their binding affinity with cortical actin. A formal description of the theory of active segregation in the asymmetric bilayer membrane with these components is presented in the SI text and Fig. S2A.

Here we find it convenient to realise this active dynamics using a Kinetic Monte Carlo simulation, whose details are in the SI. The transition rates associated with passive dynamics are determined from an energy function that depends on the short-range interaction potentials between the coarse-grained membrane components, and obey detailed balance (Fig. S2A). The transition rates associated with active dynamics correspond to two features, first, the fluctuations of the active contractile stress, represented by birth-death stochastic processes of contractile regions, and second, centripetal contractile flows of PS and TMABD determined by local gradients of the active stress, and violate detailed balance (Fig. S2B). These active transition rates depend on the following tunable parameters: (i) relative binding affinities of PS and TMABD to actin, (ii) fluctuation statistics of the active contractile stress, viz., the correlation length $\xi$ and correlation time $\tau$, (iii) magnitude of contractile stress characterized by a Péclet number Pe, or relative magnitude of active advective transport to diffusion outside, and (iv) the extent of activity or the fraction of the region influenced by active stresses. This generalizes a recent theoretical study (29, 30) of the coarse-grained dynamics of segregation of a two-component fluid driven by fluctuating active contractile stresses.

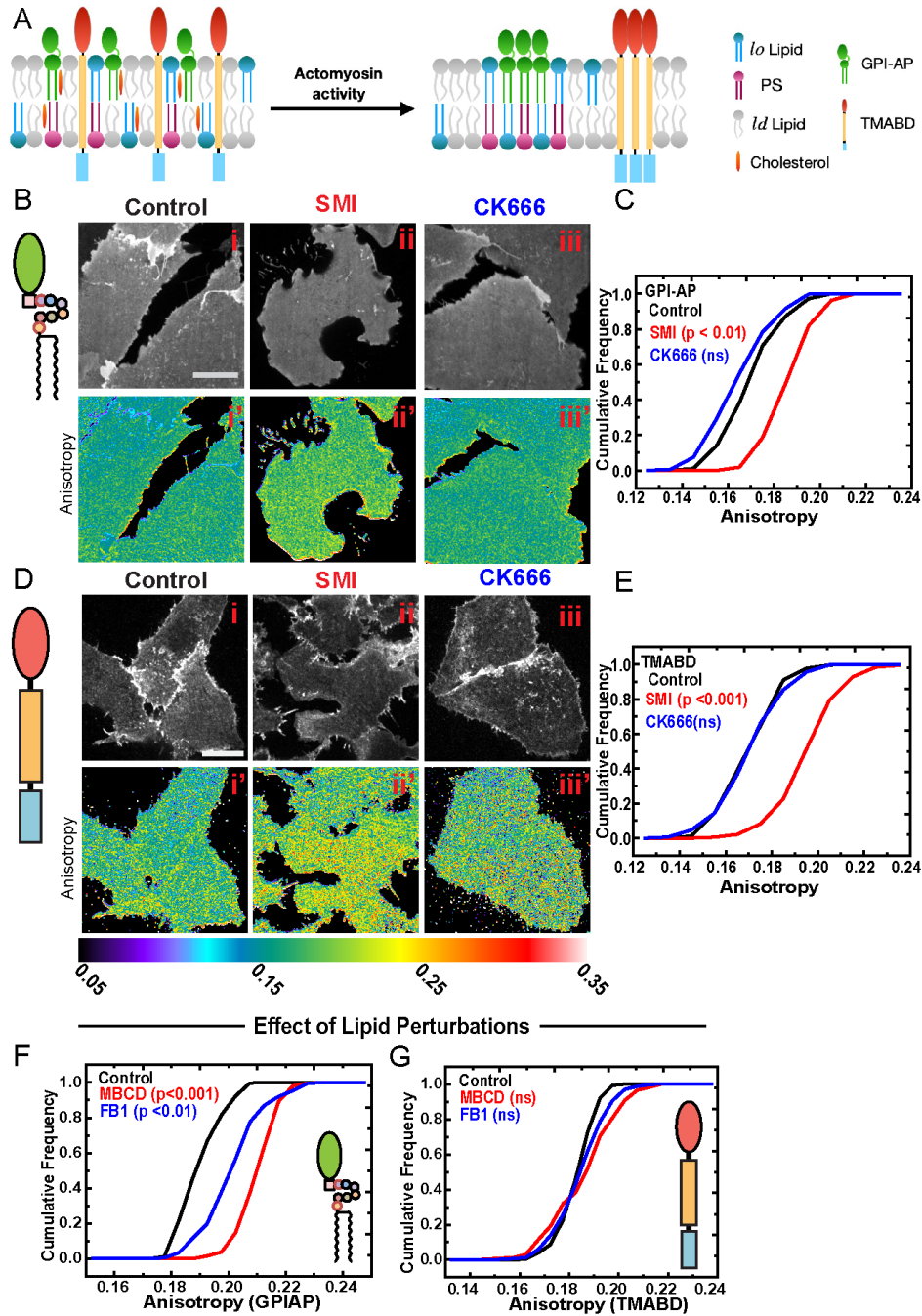

**Fig. 1. GPI-AP and TMABD nanoclusters arise from different intermolecular interactions.** A) Schematic shows actomyosin activity drives cell surface molecules like GPI-AP and TMABD into respective nanoscale clusters (on the right side). Note, while TMABD can directly couple to juxtamembrane f-actin, outer-leaflet GPI-APs can only do so via transbilayer coupling with inner leaflet PS. (B-E) Representative confocal spinning disk images (B,D) showing fluorescence intensity and anisotropy image of CHO cells expressing FR-GPI (B) or FR-TMABD (D) and labeled with PLB, either untreated (Control) or treated with pharmacological inhibitors of Formin (SMI) or Arp2/3 (CK666). Cumulative frequency plots (C, E) of anisotropy values from the indicated treatments show that formin perturbation (red line) leads to significant ($p<0.01$, KS-test) increase in anisotropy of both FR-GPI and FR-TMABD suggesting a reduction in nanoscale clustering whereas Arp2/3 perturbation (blue line) has no significant (KS-test) effects on anisotropy. (F, G) Cumulative frequency plots of anisotropy values of CHO cells expressing EGFP-GPI (F) or FR-TMABD (G; labeled with PLB) from either untreated cells (Control) or cells that were depleted of cholesterol (MBCD) or sphingomyelin (FB1). Note, only GPI-AP anisotropy (but not TMABD) increases upon these perturbations. Data were pooled from at least 15-20 cells across independent replicates (2-3) in each condition; statistical significance was tested with two sample KS-test. Scale bar for images in (B,D) : 10 $\mu m$.

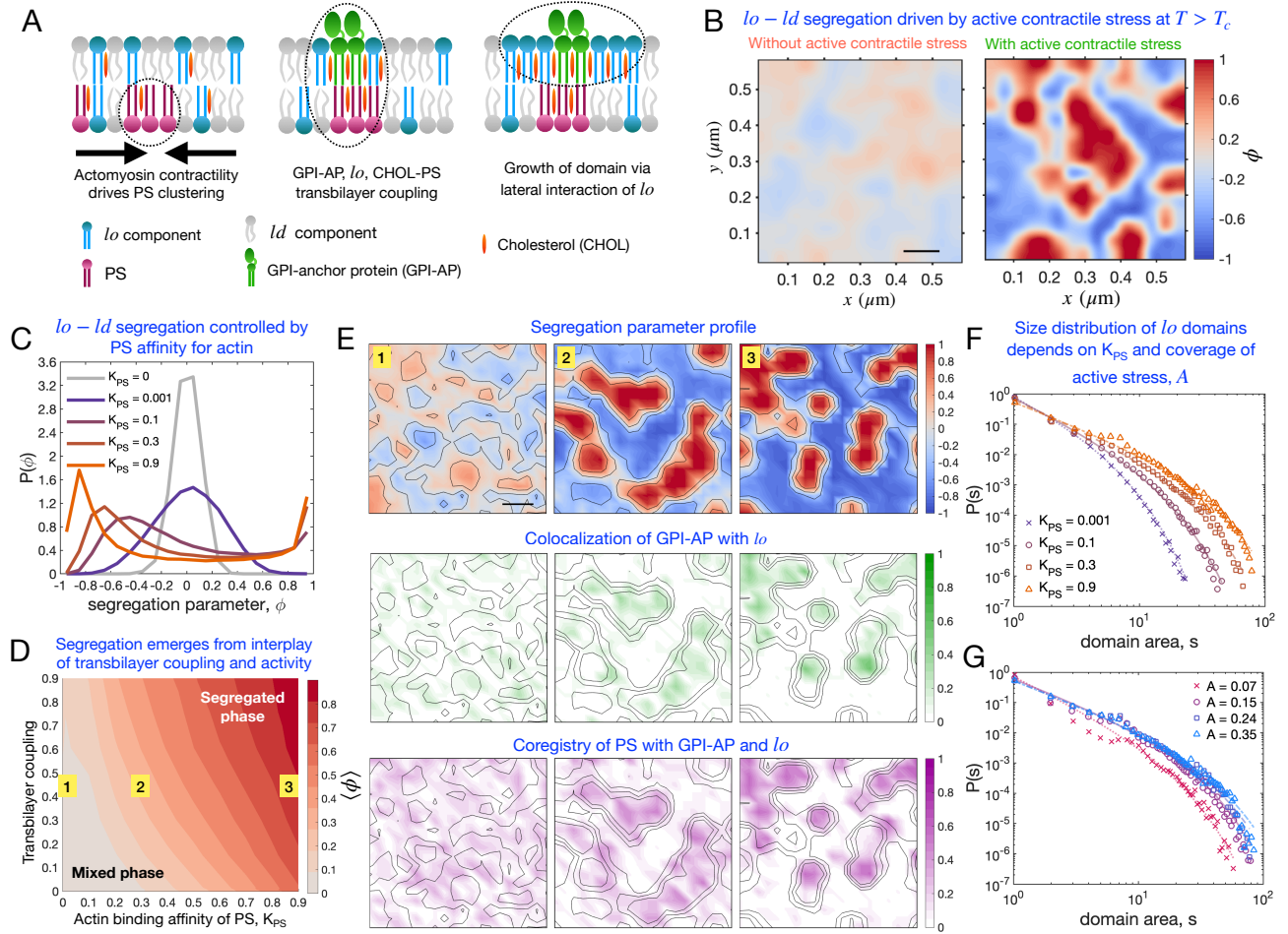

**Fig. 2. Kinetic Monte Carlo simulation of a multicomponent model of an active composite cell surface predicts nano to mesoscale lipid ordering and segregation of proteins driven by active contractile stresses generated by the actomyosin cortex.** (A) Schematic of the three principal ingredients of the multicomponent model of cell surface. For simplicity, we leave out showing the transmembrane protein TMABD, whose presence in the membrane does not make a qualitative difference to the GPI-AP associated lo-domains (see Fig. S2 for details). (B) Spatial profiles of the $lo-ld$ segregation parameter $\phi$, (right) in the presence of active contractile stresses, and (left) in their absence (all other parameters being fixed), at temperatures ($T$) higher than the critical temperature $T_c$ of equilibrium $lo-ld$ phase separation. $\phi$ is defined as: $\phi = (\rho_{lo} - \rho_{ld})/(\rho_{lo} + \rho_{ld})$, where $\rho_{lo}$ and $\rho_{ld}$ are, respectively, the local number densities of the $lo$ and $ld$ components. All simulation data correspond to $T = 1.05 T_c$, close to physiological temperature of the cell. (C) Dependence of the probability density functions (PDFs) of the segregation parameter $\phi$ on actin binding affinity of PS, $K_{PS}$ (see supplementary text for details). (D) Heat map of the order parameter, $\langle \phi \rangle$, as a function of the transbilayer coupling and $K_{PS}$. The angular brackets represent an average over multiple spatial locations and time frames. The transbilayer coupling is defined as the energy scale of PS (outer leaflet)-SM (inner leaflet) interaction $U_{SM-PS}$, expressed in units of $k_B T$, Boltzmann constant $k_B$ times the absolute temperature $T$. (E) Time averaged spatial profiles of $\phi$ (top row), number densities of GPI-AP (middle row) and PS (bottom row) at increasing $K_{PS}$ (marked by numbers on the first row in E, and in panel D). In each case, the contours of $\phi$ are also shown as a guide. All the number densities are normalized by the maximum density possible on a grid-point. (F) Distribution of the sizes of $lo$ domains (s) with increasing $K_{PS}$. (G) Distribution of the sizes of $l_o$ domains (s) with increasing value of $A$, the area fraction of active contractile stress. In F and G, the solid lines are the fits to a model distribution, $A s^{-\theta} exp[-s/s_0]$. The values of fitting parameter $A$, exponent $\theta$ and the cutoff $s_0$ are provided in Supplementary Table 1. Here the unit of s is set by the unit grid area which $0.03 \, \mu m \times 0.03 \, \mu m$. Scale bars in panel B and E are $0.1 \, \mu m$.

Our results (Fig. 2B-E) show that a combination of active stress fluctuations and strong transbilayer coupling can induce mesoscale co-segregation of $lo$-components including GPI-APs relative to $ld$-components at physiological temperatures 310 K, *which is above the equilibrium lo-ld phase segregation temperature of 293 K* determined empirically from Giant Plasma Membrane Vesicles (40). Figure 2B shows a comparison between a typical snapshot of a steady state configuration of the components in the absence of and with active contractile stresses, at temperatures above the phase segregation temperature. The extent of segregation measured by the probability distribution of the segregation order parameter $P(\phi)$ (Fig. 2C,D), and their spatial profiles (Fig. 2E), depends on the binding affinity of PS with actin (Fig. 2C) and the strength of the transbilayer coupling (Fig. 2D, E).

The panels in Fig. 2E show the active co-segregation of the $lo$-domains with GPI-AP in the upper leaflet of the membrane and their co-registry with PS in the lower leaflet. We see that the GPI-AP enriched regions are predominantly sur-rounded by the *lo*-component in the upper leaflet. The extent of segregation and co-registry increases with the affinity of PS to actin. The results on the active co-segregation of *lo*-components and GPI-AP relative to *ld*-components continue to hold when an additional component, viz. TMABD, is introduced, which only engages sterically with the other components but is driven by similar contractile stresses (Fig. S2C, D). The TMABD enriched domains are segregated from *lo*-domains, and PS (Fig. S2E).

The extent of mesoscale segregation and the mesoscale domains, depend on the fluctuating active contractile stresses, and appear at temperatures above equilibrium phase separation. This paints a picture of an *active emulsion* driven by active stirring from the cortical medium, and requiring inner leaflet PS and outer-leaflet SM with cholesterol as a glue. While, this is completely consistent with all our observations so far at the nano-scale, this description makes both qualitative and quantitative predictions about mesoscopic domains enriched in these nanoclusters, which we verify in experiments described below.

**Evidence for mesoscale organisation of GPI-AP nanoclusters.** A very clear prediction from the theoretical model is that GPI-APs nanoclusters will form mesoscale domains enriched in these entities. To explore the larger scale organisation of these nanoclusters, we re-purposed our custom-built TIRF and confocal spinning disk microscopes (31, 34) to obtain high resolution spatial anisotropy maps of the labelled proteins over a scale of 300 – 1000 nm. We identify optically resolvable domains enriched (or depleted) in these nanoclusters from spatial maps of fluorescence emission anisotropy (Fig. 3A), by thresholding the anisotropy at one standard deviation from the mean (Methods, SI; schematic in Fig. S3A). The thresholded anisotropy map of EGFP tagged-GPI (EGFP-GPI; Fig. 3A) or FR-GPI bound to fluorescent ligands (Fig. S3B), shows a nanocluster fraction in excess of $\approx 10-15\%$ of the mapped area and mean domain area $\approx 0.1~\mu m^2$ ( Control in Fig. 3B,C ), consistent with previous work (7). The dependence of GPI-AP nanoclustering on adequate levels of cholesterol and SM is consistent with their depletion severely perturbing the mesoscale domains (Fig. 3A-C). The mesoscale domains of GPI-APs are also lost when the levels of the trans-bilayer coupling lipid, PS, was perturbed in CHO cells deficient in PSS1 activity and were grown in absence of ethanolamine to reduce membrane PS levels (28) (Figs. 3D, E). As predicted, the thresholded anisotropy map of TMABD also shows mesoscale domains where the nanocluster fraction is in excess of $\approx 10\%$ of the mapped area and a mean domain area $\approx 0.1~\mu m^2$ (Figs. S3E, F, H, I). In contrast to the GPI-APs, the mesoscale organisation of TMABD is unaffected by the lipid perturbations (Figs. S3E, F). However, it is much reduced upon abrogation of the actin-binding capacity (Figs. S3H, I).

Since formin-nucleated dynamic cortical actin is necessary for building the nanoscale clusters of GPI-AP and TMABD (35), formin inhibition also leads to a significant reduction in their mean domain area (Figs. 3G, H, S3K, L; SMI), whereas Arp2/3 perturbation doesn't alter these parameters (Fig. 3G,H, S3K, L; CK666). These results confirm the qualitative predictions of the creation of mesoscale domains of nanoclusters, driven by contractile stresses in two distinct classes of molecules, GPI-APs linked via the virtue of the local lipid environment, and the TMABD to dynamic cortical actin.

**Mesoscale domain size distributions are sensitive to actin binding affinity and cortical actin dynamics.** A quantitative prediction of the theory, as elaborated previously (29), is that as a consequence of active driving, the steady state domain size distribution, $P(s)$ scales as $s^{-\theta}\exp(-s/s_0)$, a power-law with an exponential cutoff at $s=s_0$ (Figs. 2F, G). The exponent $\theta$ and the cutoff scale $s_0$ depend on the actin binding affinity, the extent of activity and correlation time $\tau$ of the active stress fluctuations (Figs. 2F, G). Using the spatial maps of the steady state fluorescence anisotropy of GPI-AP and TMABD, this domain size distribution is easily measured. We find a good fit of the mesoscale domain size distribution data to the predicted form of $P(s)$, with $\theta \approx 2$ and $s_0 \approx 0.4-0.7~\mu m^2$, for both GPI-AP (Fig. 3 C, F, I) and TMABD (Fig. S3 G, J, M ; Supplementary Table ST1).

The steady state distribution of cluster sizes still retains the above form upon perturbations of the actomyosin dynamics and the actin-binding affinity, albeit with different values of $\theta$ and $s_0$. Based on our earlier study of active segregation (29) and the predictions from the current model, $s_0$ should always decrease with any of the above perturbations. This is consistent with what we observe in our perturbation experiments (Fig. 3, S3 and Supplementary Table 1).

Together these data show that the two types of membrane components, GPI-AP and TMABD, both actively driven by cortical actin-based machinery (formin-nucleated filaments), inhabit mesoscopic membrane domains with distinct characteristics, consistent with our predictions.

**Mesoscale domains of GPI-AP are correlated across the bilayer with PS and with *lo* regions.** The active emulsion picture suggests that the mesoscale organisation of nanoclusters of outer-leaflet GPI-AP is contingent on co-registry with nanoclusters of PS at the inner-leaflet and *lo* regions (schematic in Fig. 4A, D). To study the correlation of mesoscale domains with other membrane components, we developed a two colour emission anisotropy (homo-FRET) methodology wherein we image two different components in the same cell membrane in a sequential fashion (Fig. S4A-E; detailed in SI). To examine co-registry with inner-leaflet nanoclusters of PS, we built a fusion protein of PS-binding discoidin-like C2 domain of lactadherin (LactC2) in tandem with actin-filament binding domain of Ezrin and YFP (YFP-LactC2-ABD). This fusion protein was designed to be recruited to the inner leaflet PS (41) and connect to the dynamic cortical actin machinery via its Ezrin-actin binding domain (28). LactC2-ABD-YFP forms nanoclusters at the inner-leaflet as indicated by its depolarized emission anisotropy and subsequent increase upon photo-bleaching (Fig. S4F) (25). To visualize the nanocluster-rich regions of GPI-AP and PS in the same cells we transfected YFP-LactC2-ABD encoding plasmids in GPI-anchored folate receptor (FR-GPI) expressing cells, and labeled FR-GPI with a fluorescent folate analog (pteroyl-lysyl-folate-BodipyTMR, PLB). We confirmed by sequential imaging of anisotropy of individual components, LactC2-ABD-YFP, and PLB-labelled FR-GPI co-existed in highly correlated regions (Fig. S4D, E, G). Analysis of the correlation of the spatial anisotropy maps of the outer and inner-leaflet localized proteins (Fig. 4B) showed that the cluster-rich (low emission anisotropy) regions of the LactC2ez are strongly correlated with the cluster rich regions of GPI-AP (Pearson coeff.= 0.6, p < 0.01; Fig. 4C), but only moderately with TMABD (Pearson coeff.= 0.28, p <0.01; Fig. S4H, I). Together with our previous findings (28), this shows that the actively generated mesoscale organisation of GPI-AP nanoclusters occurs at sites of enrichment of PS clusters in the inner-leaflet, mediated by transbilayer coupling, as predicted by the theory.

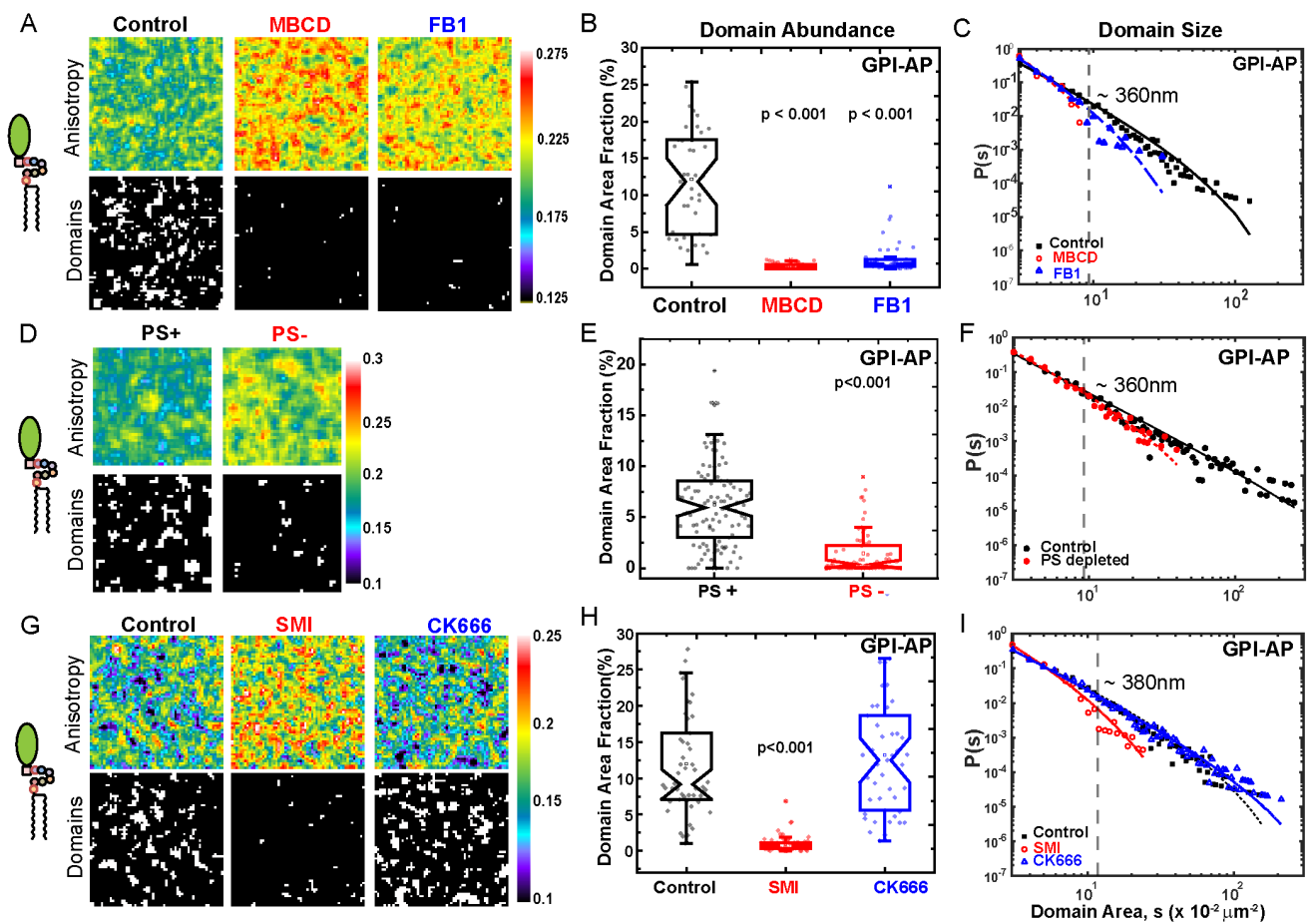

**Fig. 3. Mesoscale organisation of GPI-AP nanoclusters.** (A) Boxes represent $6\ \mu m \times 6\ \mu m$ patches of anisotropy map (top) and corresponding thresholded binary maps (below, showing mesoscale domains enriched in nanoclusters) taken from CHO cells expressing EGFP-GPI which were untreated (Control) or depleted of cholesterol (MBCD) or sphingomyelin (FB1). Notched box-plots (B) of Domain Area fraction obtained from several images like (A) show mesoscale GPI-AP domains are significantly reduced upon these lipid perturbations (p<0.001 One-way ANOVA with Tukey mean-comparison). (C) Distribution of domain area in the indicated conditions; solid lines represent the corresponding fits to a model distribution, $As^{-\theta}exp[-s/s_0]$. The characteristic area of GPI-AP domains in Control scenario is estimated to be $0.1\ \mu m^2$ (dashed vertical line) which translates to a domain diameter of $\approx 360\ nm$, while both the lipid perturbations mark a shift towards smaller domain area as indicated decrease in the frequency P(s) of larger domains ($s \geq 0.1\ \mu m^2$). (D-F) Phosphatidylserine (PS) deficient PSA3 cells transiently expressing EGFP-GPI were grown in PS+ or PS- conditions. Anisotropy and binary domain maps (D; $6\ \mu m \times 6\ \mu m$ boxes) show that PS depletion led to a reduction in domain area fraction (E, p< 0.001, Mann-Whitney test), and Domain area distributions (F) show a shift toward smaller domain areas for PS depleted cells (red circles and dotted fit line). (G-I) Boxes (G) show $6\ \mu m \times 6\ \mu m$ anisotropy and binarized domain maps of PLB-labeled FR-GPI obtained from untreated cells (Control), or cells treated with inhibitors of formin (SMI) or Arp2/3 (CK66). Notched box-plot (H) of domain area fraction shows that only formin perturbation leads to loss of these domains and led to a marked shift to smaller $s$ in domain area distributions (I). Statistical significance was determined at p < 0.001 using one-way ANOVA with Tukey mean-comparison. At least 30-60 thresholded binary maps were pooled from 10-15 cells across 2 independent replicates for the analyses of each condition.

Next, we sought to understand the local lipid order of these cluster rich (or sparse) mesoscopic domains of FR-GPI. To do so we map local membrane order using the polarity sensing lipid probe Laurdan, and quantify it by measuring Generalized Polarization (GP) (42–44). GP values are low(er) in *ld*-regions and high(er) in *lo*-regions of the membrane. While the dynamic range of the Laurdan GP values in CHO cells is not very large, as has been noticed in earlier cellular studies (45) and in cell derived Giant Plasma Membrane Vesicles (46), it is sufficient to identify regions enriched in *lo* or *ld* components.

We obtained sequential spatial maps of Laurdan GP and emission anisotropy (homo-FRET) of GPI-AP using a spinning disk confocal microscope on a uniformly labelled basal membrane (Fig. 4E, S5A-D). We measured the anisotropy coarse-grained over $4 \times 4$ pixels (pixel:102 nm) and determined the cluster-rich (cluster-sparse) regions together with the corresponding Laurdan GP from membrane patches across multiple cells (Fig. 4E). We found that GPI-AP nanocluster-rich domains have a higher Laurdan GP compared to GPI-AP nanocluster-poor domains (Fig. 4F, G). On the other hand, the TMABD cluster-rich (or sparse) regions have very little difference in their local membrane order (Fig. S5E-G). These results suggest that mesoscale domains enriched in GPI-AP nanoclusters have a specific *lo*-lipid environment, while the TMABD nanoclusters are agnostic to its lipid environment, exactly as predicted by our theory.

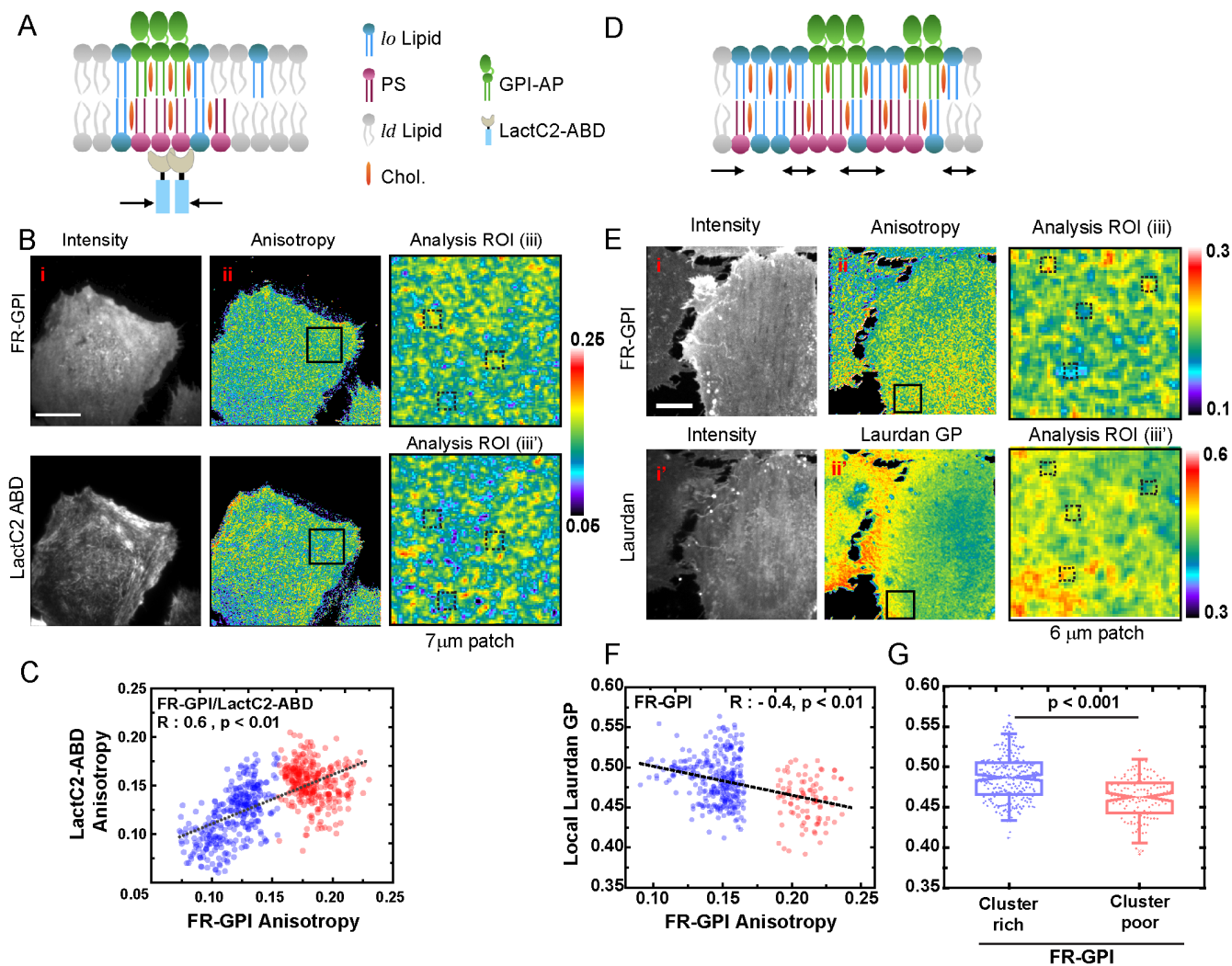

**Fig. 4. Mesoscale domains of GPI anchored protein clusters are correlated with inner-leaflet PS cluster-rich and *lo* domains.** (A) Schematic shows transbilayer coupling between the GPI-AP and inner-leaflet PS. LactC2-ABD which acts as a synthetic linker between the inner-leaflet PS and juxtamembrane actin flows (black arrows). Representative TIRF images of fluorescence intensity and anisotropy of PLB-labeled FR-GPI (B, top panels), transfected with YFP-LactC2-ABD (B; lower panels). Matched ROIs of $7 \times 7 \mu m$ sized patches (black boxes in B ii) were chosen from the anisotropy images of PLB-labeled FR-GPI and LactC2-ABD (iii, iii'). Mesoscale domain sized ROIs ($\approx 400 \times 400$ nm) are drawn around cluster-rich and cluster-poor hotspots of FR-GPI (B iii, iii'). The anisotropy values of FR-GPI and LactC2-ABD graphed as scatter plots (C) exhibit significant positive correlation (Pearson's R = 0.6, p < 0.001). Individual scatter plots report 630 ROIs collected from cells (n=29 cells) pooled from two independent replicates. (D-G) Schematic (D) shows how individual nanoscale GPI-AP cluster co-habit a larger mesoscopic patch of lo-like membrane driven by actin flows to form 'active emulsions'. Representative high-resolution confocal fluorescence and anisotropy images of CHO cells expressing FR-GPI labeled with PLB (E, top panel) and co-labelled with membrane order probe, Laurdan (E, bottom panel). Identical $7 \mu m \times 7 \mu m$ sized boxes (in black) were chosen from the anisotropy and Laurdan GP map and shown in Analysis ROI (E; iii, iii', respectively). Scatter plot (F) of local Laurdan GP and anisotropy computed from corresponding small regions (400 x 400 nm; dashed boxes) obtained from Analysis ROI for corresponding FR-GPI anisotropy (iii) and Laurdan GP (iii') maps. Note Laurdan GP and anisotropy of PLB-labeled FR-GPI shows negative correlation (R= -0.4, p < 0.01). (G) Box plots show cluster-rich domains of GPI-AP have significantly higher local Laurdan GP (p< 0.001, two sample t-test). Dataset (F,G) report on at least 500 ROIs collected from cells ( n = 15 cells) pooled from two independent replicates. Scale bar $10\mu m$.

**Lateral membrane ordering is contingent on dynamic cortical actin and transbilayer coupling.** An important claim of the active emulsion description is that the mesoscale *lo*-region observed at physiological temperatures is not a consequence of a thermodynamic phase transition, but rather driven by active contractile stresses arising from cortical actomyosin that act via the inner-leaflet PS to maintain a strong transbilayer coupling to GPI-AP. To test this crucial prediction, we study how local membrane ordering, character-ized by Laurdan GP, gets affected upon perturbing the actin-nucleator formin, actin (de)polymerization, and inner-leaflet PS.

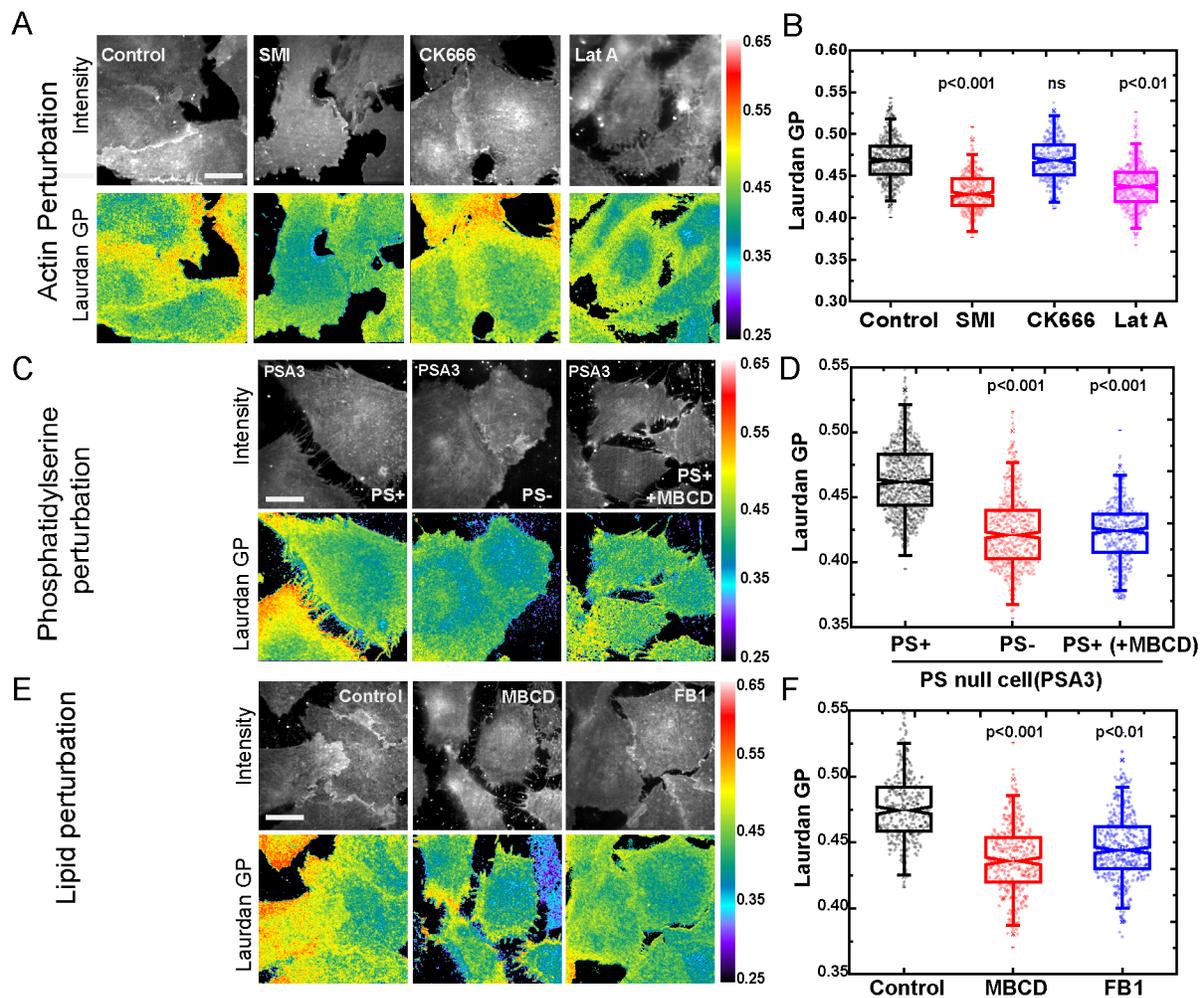

**Fig. 5. Molecular components driving the formation of ordered domains.** (A, B) Representative confocal images (A) show Laurdan total intensity (Intensity) and GP map of untreated CHO cells (Control) or cells treated with different perturbation of actin cytoskeleton. (B) Box-plot shows the effects of different actin perturbations on Laurdan GP. Perturbation of actin polymerization (Lat A; pink) or formin (SMI; red) but not Arp2/3(CK666; blue) lead to significant reduction of Laurdan GP compared to control (black). (C, D) Laurdan Intensity and GP images (C) of the PS biosynthetic mutant CHO cells (PSA3) grown in presence (PS+) or absence (PS-) of ethanolamine. Box-plot (D) shows that Laurdan GP of PS- (red) cell membrane is reduced significantly compared to PS+ (black), suggesting a reduction of steady state membrane ordering. Cholesterol depletion of PS+ cells by MBCD (C; D, blue) leads to similar loss of membrane ordering. (E, F) Representative confocal images (E) of CHO cells show Laurdan intensity and GP maps of untreated (E, Control), or cells depleted of cholesterol (MBCD) or sphingomyelin (FB1). (F) Box-plot of Laurdan GP shows the reduction of membrane ordering in cholesterol (red) and sphingomyelin (blue) depleted cells compared to control (black). Dataset for each condition was collected from at least 300-500 regions (individual dots in box plots) from 15-20 cells; Scale bar for images in A,C,E $10\mu m$. Statistical signficance was defined at $p< 0.001$ using One-way ANOVA with Tukey's mean comparison test.

Perturbations of formin (SMIFH2, $25\mu M$, 2 hr) but not Arp2/3 (CK666, $50\mu M$, 2 hr) led to a reduction in GP values on CHO cells as measured using two different probes for membrane ordering, *viz.* Laurdan (Fig. 5) and the outer-leaflet specific Nile Red analogue NR12S (47, 48) (Fig. S6). Cells treated with formin inhibitor, SMIFH2 (Fig. 5A; SMI), and a significant reduction in steady state Laurdan GP and NR12S ratios (Fig. 5 A, B and Fig S6 A,B; SMI) while treatment with Arp2/3 inhibitor does not affect the GP or NR12S ratios compared to control (Fig. 5 A, B and Fig. S6 A,B; CK666). Finally, latrunculin A mediated inhibition of actin polymerization in general (Fig. 5A,B Lat A) also lowers GP value. Note that for the NR12S probe, the shift in GP values is small, but statistically significant (Fig. S6 A,B; Lat A). Laurdan GP measured from PS-deficient cells (PS-) shows a clear drop in steady-state membrane order in comparison to cells grown in presence of ethanolamine (PS+ in Fig. 5C,D). The extent of decrease in Laurdan GP in PS- cells are comparable to those obtained from PS+ cells depleted of cholesterol (PS+ MBCD, Fig. 5 C,D), buttressing the critical role of innerleaflet PS in maintaining local membrane order. Not surprisingly, the depletion of membrane cholesterol (MBCD) and outer-leaflet sphingomyelin (FB1) in wild type cells also show marked reduction in the membrane ordering (Fig. 5E, F). These observations verify a critical prediction of the theory, namely that the concerted involvement of dynamic cortical actin and transbilayer interaction via inner-leaflet PS drives the local *lo*-membrane order that accompanies the mesoscale organisation of GPI-AP nanoclusters.

## Discussion

Several studies lend support to the picture of the cell surface as a multi-component composite- an asymmetric bilayer membrane sandwiched between a thin cortical layer of actomyosin and an extracellular matrix (49). The membrane itself has rich compositional diversity, consisting of a variety of lipids and proteins. In this context many transmembrane protein receptors have evolved intracellular domains allowing them to interact with the cortical cytoskeleton either directly or via recruiting actin-binding domains such as Ezrin, ankyrin or alpha catenin (50, 51). On the other hand, proteins and lipids localized to the outer-leaflet of the plasma membrane, such as the lipid-tethered GPI-APs and glycolipids, also interact with cortical actin, albeit in an indirect manner. Their interaction is mediated via the creation of a specific *lo*-lipid environment at nanoscales, involving acyl-chain transbilayer coupling with inner-leaflet phosphatidylserine (PS) and the stabilizing influence of local cholesterol and sphingomyelin (12, 28).

The active composite membrane model elaborated here, describes the dynamical response of GPI-APs (and possibly glycolipids) at the outer leaflet as a consequence of being internally driven by active actomyosin-dependent stresses. As we show here, it predicts that these nanoclusters are organised in mesoscale domains that exhibit *lo* characteristics. Validating these predictions we find mesoscopic domains enriched in clusters whose size is $\approx 350-500$ nm, consistent with earlier high resolution studies (7, 33, 52). This lateral mesoscale organisation of nanoclusters is contingent on (i) transbilayer coupling to inner-leaflet PS, (ii) the binding of PS to the active machinery and consequently subject to active contractile stresses, and (iii) the presence of *lo*-lipid components such as SM and Cholesterol, and as a consequence manifests at physiological temperatures, *which is significantly higher than the thermodynamic lo-ld phase transition*. As confirmed in experiments the *lo*-domains are dependent on a dynamic actin cortex comprising formin-nucleated actin filaments.

Figure 6 shows a schematic of the emergence of this mesoscale organization driven by contractile active stresses, transbilayer coupling via PS and lateral lipidic interactions. This pre-existing lateral segregation on the flat asymmetric bilayer cell membrane is aptly described as an *active emulsion* formed by local non-equilibrium stirring of the composite. We expect such active emulsions to display anomalous fluctuations in their non-equilibrium steady state (24, 30). This description differs from passive (micro)emulsion models, such as those that invoke preferential wetting as in the lipid-shell model (5), or that impute line-actants (53, 54), or depend on the coupling of membrane shape fluctuations to local lipid composition (55, 56), in two fundamental aspects. First, there is a new molecular organiser of local plasma membrane composition, namely cortical actomyosin and its linkage to the lower leaflet lipid PS, along with attendant lateral interactions. Second, it is dependent on novel forces for driving local organisation, namely non-equilibrium active contractile stresses and passive transbilayer interactions. It is the combination of active internal forces driven by a molecular mechanism involving actomyosin, and specific passive interactions, including transbilayer interaction with PS and lateral interaction with *lo*-components, that outcompetes the entropy of mixing and drives lateral organisation of lipids at the nano and mesoscales in the plasma membrane at physiological temperatures.

Concurrently, the lateral active organisation of transmembrane proteins such as TMABD that bind directly to actin, appears to be agnostic of the underlying membrane composition. Their mesoscale organisation arises from a combination of active internal forces driven by actomyosin and passive protein-protein interactions. This may provide an explanation for the mesoscale domains of several actin-interacting endogenous proteins which also exhibit *cis*-interactions in their ecto-domains such as CD44 and E-Cadherin (10, 11, 57). In the event that there are covalent or non-covalent interactions of these proteins with specific lipids that prefer *lo* or *ld*-phases, we propose that such actively driven protein clusters will also be correlated with specific lipid environments, dictated by these interactions.

The *active emulsion* picture elucidated here has parallels in the creation and maintenance of liquid-liquid phase separated assemblies (58, 59), otherwise known as biomolecular condensates (60). Apart from purely thermodynamic agencies driving such biomolecular aggregation (58–60), it is likely that activity, arising from either nonequilibrium chemical reactions (61) or active cytoskeletal elements (as discussed here), could provide the driving force for the formation and maintenance of such mesoscale emulsions.

The pre-existing dynamic organisation of the plasma membrane described here is maintained out-of-equilibrium by constitutive active processes. This active non-equilibrium steady state sets up the plasma membrane to respond sensitively to stimuli, such as in *outside-in and inside-out* cell surface signalling (18). It is therefore no surprise that many membrane signalling receptors have evolved intricate direct or indirect molecular mechanisms linking them to the cytoskeleton (10, 11, 57, 62).

For instance, the hierarchical architecture of having mesoscale domains of membrane proteins, built from dynamic nanoscale motifs, allows for a rapid remodelling of an activated zone (13, 18, 22). The strength and nature of the coupling between the actomyosin machinery and membrane components could be locally regulated, to build larger platforms or disassemble existing ones. A striking example of this, is the rapid expansion of an activated zone of GPI-AP nanoclusters induced during local integrin activation and cell substrate adhesion, with consequences for cell physiology and receptor function (35).

Finally, there are two immediate and important consequences of the induced and pre-existing active organisation of membrane components. One is the central involvement of actomyosin in nano and meso scale organisation. This allows the cell to explore mechano-chemical mechanisms for the local control of cell membrane organisation, and thereby mechano-chemical signal processing and feedback (63, 64). The sec-ond is the key finding of differential segregation of membrane components driven by active contractile stresses. This must direct the focus to uncovering the molecular basis of active contractile force generation at the cortex - their chemistry and interaction partners, force characteristics, spatial organisation and turnover.

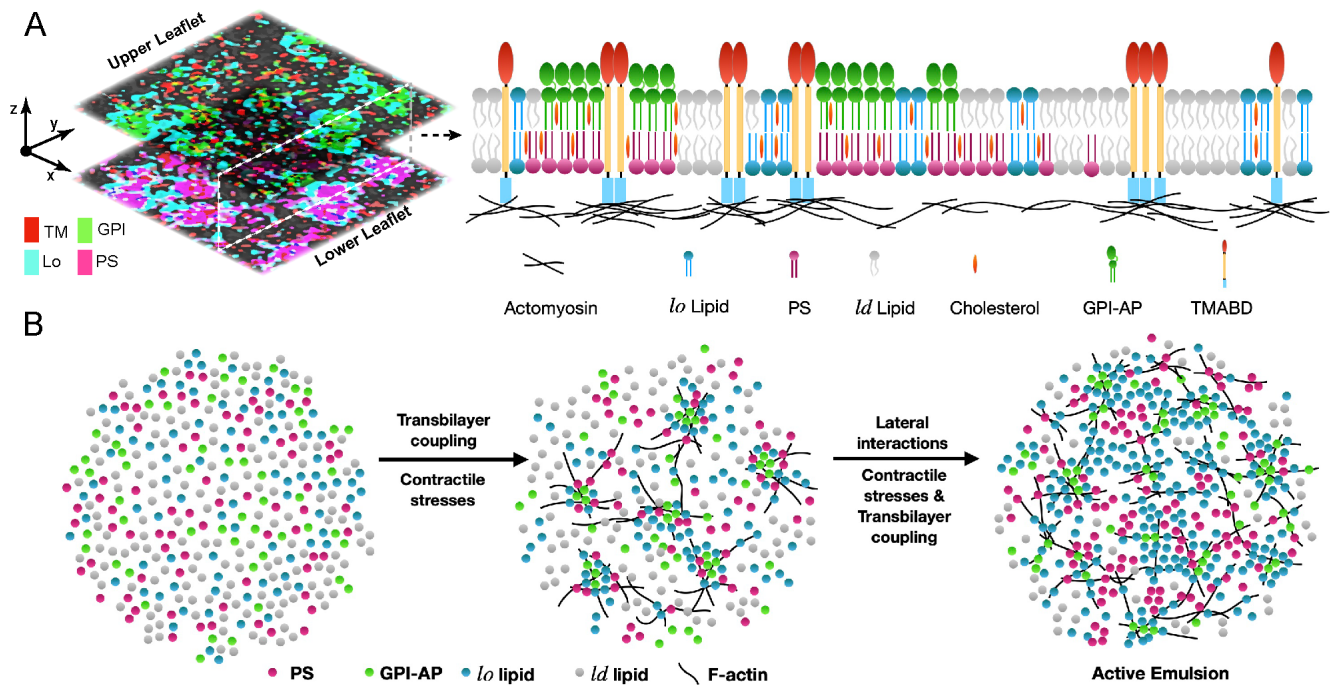

**Fig. 6. Active emulsions: Emergent mesoscale domains at the cell surface.** (A) Image (left) shows a spatially averaged map of upper and lower-leaflet components obtained from a snapshot of simulations described in Figure 2. An axial-slice (top right) of the cell membrane schematises the organisation of the resultant patches of GPI-AP clusters enriched in lo-membrane components and juxtamembrane actin interactions at the inner-leaflet. (B) Model for the emergence of 'active emulsions'. Transbilayer coupling and contractile stresses organise GPI-APs and lo-lipids present in a uniformly mixed bilayer (left) to form individual nanoscale cluster (middle), which in turn sets up the platforms to facilitate local lipid-lipid lateral interactions leading to the emergence of ordered mesoscopic domains or 'Active Emulsion'

## Materials and Methods

**Cell lines, labeling and perturbation.** Chinese hamster ovary (CHO) cell lines were maintained in Ham's F12 media (Hi Media, India) supplemented with (10%) fetal bovine serum (FBS, Gibco, USA). Plasmids were either expressed stably (FR-GPI, FR-TMABD, EGFP-GPI) or transiently transfected (LactC2-ABD-YFP) prior to imaging. PSA3 cell line for phosphatidylserine (PS) perturbation experiments were maintained in Ham's F12 medium in the presence of 10 mM ethanolamine and 10% FBS. N$\alpha$-pteroyl-N$\varepsilon$-BodipyTMR-L-lysine (PLB) was used to label folate-receptor (FR-GPI and FR-TMABD) expressing cells at saturating concentration ($\sim$ 400 nM) for 5 mins at 37°C. Imaging were carried out in a HEPES (20 mM) based buffer supplemented with 2 mg/ml glucose (details in SI methods).

Perturbations were carried out as following: (1) Cholesterol: Methyl-$\beta$-cyclodextrin (MBCD); 10 mM at 37°C for 30 minutes. (2) Sphingomyelin: Fumonisin B1 (FB1); 40 $\mu$g/ml for 72 hours in growth culture media. (3) PS depletion: PSA3 cells were grown in media devoid of ethanolamine and 10% dialyzed FBS for 36-48 hours. (4) F-actin: Latrunculin A (LatA); 2 $\mu$M for 10-15 min at 37°C. (5) Formin: SMIFH2; 25 $\mu$M for 2 hr at 37°C. (6) Arp2/3: CK-666; 50 $\mu$M for 2 hr at 37°C. Unless noted otherwise, the drugs against F-actin, formin and Arp2/3 were maintained in the imaging buffer during microscopy.

**Steady-state anisotropy measurements.** Steady state anisotropy experiments were carried out as described earlier (31, 34) using the following high resolution (100x Objective, 1.45 NA) imaging platforms: (1) Yokogawa CSU-22 spinning disk (SD) confocal microscope (Andor Technologies, Northern Ireland); (with a sampling pixel size of 100 nm) or 2) Nikon TIRF microscope (with a pixel size of 75 nm). Image processing, analysis, and quantification were performed using Metamorph 7.0 (Molecular Devices Corporation, CA, USA), MATLAB (Mathworks, MA, USA) and ImageJ (NIH, USA) as described earlier and detailed in SI.

**Membrane ordering measurements.** Membrane ordering measurements were primarily carried out by measuring the general polarization (GP) of Laurdan (6-lauryl-2-dimethylamino-napthalene) at the cell surface. Cells were co-labeled with Laurdan (at 10 $\mu$M) and folate-analog PLB (at 400 nM) at 37°C for 5 minutes. Laurdan labeled cells were imaged at 100x (1.45 NA) on a confocal spinning disk microscope using 405 nm laser line and emission fluorescence was recorded at the two spectral channels 410-460 nm (Ch1) and 470-530 nm (Ch2). Laurdan GP calculation and standards were implemented as described earlier and detailed in SI (32, 43).

In some experiments, Nile Red based membrane dye

(NR12S) blue-red ratiometric imaging provide an alternate measure of lipid order (48) as detailed in the SI.

**Data Analysis.** *Quantification of domain abundance and size distributions.* Pixel anisotropy distribution obtained from several (20) 6-10 $\mu m$ square patches of spatial anisotropy maps of a probe (say FR-GPI in control conditions) were used determine the mean and standard deviation (SD) and set the threshold (Mean–SD). Mesoscopic domains were demarcated by thresholding and binarizing spatial anisotropy maps. For comparisons between control and perturbations, same threshold determined from 'control' was used uniformly across all conditions. These binarized maps were quantified using 'Analyze particles' routine of ImageJ with a size criterion of at least 5 pixels (pixel size is $0.1\mu m$ or $0.01\mu m^2$) to obtain domain area fraction (for each patch) and individual domain area.

*Quantitative relation between clustering (of two probes) and with lipid order.* A systematic correlation-based approach was implemented to compute the relation between the clustering (from anisotropy values) for two probes or that between clustering and lipid order (Laurdan GP). Briefly (see details in SI), anisotropy rich and poor regions were manually identified (say for FR-GPI; probe 1) and demarcated with a region-of-interest (ROI) box surrounding the domains. The size of the box (0.4x0.4$\mu m$; 0.16 $\mu m^2$) was guided by the mesoscopic domain sizes we determined earlier and several ROIs (at least 300-500) were marked across the entire dataset (at least 15 cells across 2-3 independent replicates). These ROIs were then used to calculate anisotropy values of either both the probes (say FR-GPI and LactC2-ABD) or anisotropy of a given probe (FR-GPI) and it's corresponding GP values from Laurdan maps. The data was displayed as scatter plots to show the qualitative trend of correlation and compute the Pearson's correlation coefficient. Further details of the imaging data analysis can be found in SI.

**Theory and Simulation.** We provide a formal description of the theory of active phase segregation in a multicomponent asymmetric bilayer membrane in the supplementary text and Fig. S2, which includes the details of the components, the ingredients of the theory (active contractile stresses, transbilayer coupling and lateral interactions) and a detailed account of the methodology used for Kinetic Monte-Carlo Simulation.

**ACKNOWLEDGEMENTS**
We thank members of MR and SM laboratories for stimulating discussions and critical reading of the text. We thank Andrey Klymchenko (University of Strasbourg) for the kind gift of Nile Red Dye (NR12S). SS acknowledges fellowship supports from NCBS (TIFR) and American Heart Association (18POST33990156). AD acknowledges support from the Centre for Theoretical Biological Physics and the Discovery Cluster at Northeastern University; AAA was supported by N-PDF fellowship (SERB-DST); CP and PS acknowledge graduate fellowship from NCBS (TIFR); MR and AD acknowledge support from the Simons Foundation (Grant No. 287975). MR and SM are J.C. Bose Fellows (Department of Science and Technology, Government of India) and SM acknowledges support from Human Frontier Science Program Grant (RGP0027/2012) and DBT-Wellcome Trust India Alliance Margadarshi fellowship (IA/M/15/1/502018). Authors acknowledge support from NCBS Central Imaging and Flow Facility (CIFF) and the computing facilities.
The authors declare no competing financial interests.
**Author Contributions:** Conceptualisation: SS, AD, SM, MR; Investigation: SS, CP, AAA, PS; Formal Analysis: SS, AD, CP; Theory and Simulations: AD, MR; Visualisation and Figures: SS, AD; Writing and revising: SS, AD, MR and SM with inputs from all authors.

Supplementary Information for

# Active emulsions in living cell membranes driven by contractile stresses and transbilayer coupling

Suvrajit Saha, Amit Das, Chandrima Patra, Anupama Ambika Anilkumar, Parijat Sil, Satyajit Mayor and Madan Rao

## Supplementary Note 1: Coarse grained description of a multicomponent asymmetric bilayer driven by PS-actomyosin interaction, active contractile stresses and transbilayer coupling

**A. Multi-component asymmetric bilayer membrane.** A minimal description of the nano-to-meso scale organization of lipids and GPI-AP in the multicomponent asymmetric bilayer, driven by active contractile stresses and transbilayer coupling requires at least 5-components, $\alpha = 1, \ldots, 5$ (Fig. S2a)

1. GPI-AP in the upper-leaflet,

2. *lo*-component in the upper-leaflet, collectively comprising SM, cholesterol, ...,

3. PS in the lower-leaflet, together with cholesterol and other long-chain saturated lipids such as PIP2 (other *lo* component in Fig. S2a),

4. *ld*-component in the upper leaflet, collectively comprising short chain, unsaturated lipids.

5. *ld*-component in the lower leaflet, collectively comprising short chain, unsaturated lipids.

In the cases where we compare the organization of GPI-AP with a generic transmembrane protein that can bind to cortical actin, we include one more component, $\alpha = 6$,

6. transmembrane protein TMABD straddling both leaflets,

In our coarse grained description we represent all molecules as spheres, that reside either on the upper or lower leaflet of the bilayer membrane, or straddling the two leaflets (partitioned according to Rules 1-6 above). The two leaflets of the bilayer are treated as two parallel 2-dimensional planes, separated by a distance $\sigma$, and lateral dimension $L \times L$. The total number of molecules in the two planes $N$, is high enough so that there are no large empty spaces in either plane.
Components that reside on the same leaflet of the bilayer membrane interact via a Lenard-Jones potential of the form,

$$V(X_i^\alpha, X_j^\beta) = \infty, \text{ if } r_{ij}^{\alpha\beta} \leq \sigma_{\alpha\beta}$$
$$= -J_{\alpha\beta}\left(\sigma_{\alpha\beta}/r_{ij}^{\alpha\beta}\right)^6, \text{ if } r_{ij}^{\alpha\beta} > \sigma_{\alpha\beta} \quad (1)$$

where $i$-th molecule of components $\alpha$ interacts with the $j$-th molecule of component $\beta$, with an attractive interaction strength $J_{\alpha\beta}$, expressed in units of thermal energy $k_B T$ ($k_B$ is the Boltzmann constant and $T$ is the absolute temperature). Here, $\sigma_{\alpha\beta}$ represents the hard-core length scale and $r_{ij}^{\alpha\beta} = |X_i^\alpha - X_j^\beta|$ is the distance between the two molecules. For simplicity, we take $\sigma_{\alpha\beta} = \sigma$, for all $\alpha, \beta$. Note that with the above ranking of the components, the only cross-term that is nonzero is $J_{12} > 0$, the rest are 0.

Components that reside on separate leaflets of the bilayer membrane do not have any steric interaction, and only interact via an attractive Lenard-Jones potential,

$$V_\pm(X_i^\alpha, X_j^\beta) = -J_{\alpha\beta}^\pm (\sigma_{\alpha\beta}/r_{ij}^{\alpha\beta})^6 \quad (2)$$

For simplicity, we take $\sigma_{\alpha\beta} = \sigma$, for all $\alpha, \beta$. With the above ranking of the components, only $J_{13}^\pm = J_{23}^\pm > 0$, the rest are 0.

The TMABD being transmembrane has just steric interactions with components on either leaflet.

**B. Active contractile stresses from actomyosin cortex.** The cortical actomyosin adjoining the lower leaflet of the bilayer applies stochastic contractile stresses on the membrane components that bind to it - these are the lower leaflet PS ($\alpha = 3$) and the transbilayer protein TMABD ($\alpha = 6$), with binding affinities given by, $K_{PS}$ and $K_{TM}$, respectively.

Following (1–3) we take the statistics of the contractile stresses to be telegraphic - a birth-death process - that is exponentially correlated in time, with a mean lifetime $\tau$. These contractile stresses are strongly correlated over a spatial scale $\xi$ and uncorrelated otherwise. Let us denote as $\{\Omega_{\mathbf{x}_a}\}$, the region of radius $\xi$ centred around $\mathbf{x}_a, a = 1, \ldots n$, within which the active stresses are nonzero and contractile,

$$\Sigma(x,t) = -\Sigma_0 \, \hat{\mathbf{r}}_a, \text{ if } x \in \Omega_{\mathbf{x}_a}$$
$$= 0, \text{ if } x \notin \Omega_{\mathbf{x}_a} \quad (3)$$

where $\hat{\mathbf{r}}_a$ is the unit radial vector from the centre of the domain $\Omega_{\mathbf{x}_a}$. We work in an ensemble where the number of such active stress events $n$ are held fixed (alternatively, we can specify the birth-death rates which fixes the mean $\langle n \rangle$). The coverage of nonzero active stress is $A = n\pi\xi^2/L^2$.

## Supplementary Note 2: Dynamics of segregation : Kinetic Monte-Carlo Simulation

The dynamics of the membrane components, subject to both equilibrium and active forces, are described in terms of a Master equation for the time evolution of the probability distribution, $P(\{X_i^\alpha\}, \{\Omega_{\mathbf{x}_a}\}, t)$. We solve the Master equation using a kinetic Monte Carlo approach, where we specify the updates for the positions $\{X_i^\alpha\}$ of the membrane components and placement $\{\Omega_{\mathbf{x}_a}\}$ of the active stress events.

For numerical convenience, the membrane components on the two planes are restricted to lie on a square lattice with lattice spacing $\sigma$ (the square lattice in the two planes are in complete registry). The configuration in each plane is updated by attempting an exchange of the positions of any chosen pair of neighbouring particles, $\{X_i^\alpha\}, \{X_j^\beta\}$. These position exchanges are determined both by equilibrium pair-potentials (Eqs. 1, 2) and by forces arising from the local contractile stress if the particles (PS and TMABD) are in the region $\{\Omega_{\mathbf{x}_a}\}$ (Eq. 3).

The details of the transition rates associated with the exchange moves are described below and in Fig. 1b. Note that one time step of this kinetic Monte-Carlo scheme, $\Delta t$ is given by $N$ attempts at position exchanges ($N$ is the total number of particles in the two planes). We choose $\Delta t$, so that $\sigma^2/(4\Delta t)$ is the typical diffusion coefficient of a protein on the membrane.

Every lattice site on the upper plane is occupied by one of the components $\alpha = 1, 2, 4$, similarly every lattice site on the lower plane is occupied by one of $\alpha = 3, 5$. The transmembrane protein $\alpha = 6$, when included, will straddle both the planes.

**A. Equilibrium exchange moves.** Choosing a neighbouring particle pair in the same plane - $i$ (of component $\alpha$) and $j$ (of component $\beta$) - with uniform probability, we attempt an exchange with a probability,

$$\begin{aligned} \omega &= 1, \quad \text{if } \Delta \mathcal{E} \leq 0 \\ &= e^{-\Delta \mathcal{E}/k_B T}, \quad \text{if } \Delta \mathcal{E} > 0 \end{aligned} \quad (4)$$

where $\Delta\mathcal{E}$ is the difference in energy before and after the exchange, and is obtained from the inter-particle potentials Eqs. 1, 2. This is the usual Metropolis transition probability and obeys detailed balance (4).

**B. Active advection moves.** This applies only to PS ($\alpha = 3$) and TMABD ($\alpha = 6$) when included. When these particles are within the contractile regions $\Omega_{\mathbf{x}_a}$ and bound to it with affinities, $K_{PS}$ and $K_{TM}$, respectively, they move preferentially towards the centre $\mathbf{x}_a$ with a radial velocity proportional to $\Sigma_0$ (Eq. 3), by attempting a series of exchange moves. This detailed balance violating move is shown in Fig. S2b - for details of the implementation see (3).

**C. Choice of Parameters and simulation units.** In the simulations we vary the parameters: (i) total coverage of the contractile regions where active stress is nonzero, $A$; (ii) life time of the active stress events $\tau$; (iii) binding affinities, $K_{PS}$ and $K_{TM}$; and (iv) transbilayer coupling tuned via $J_{23}$. The rest of the parameters of the model, such as overall composition of the components and temperature $T$, are held fixed. The qualitative results we obtain are independent of these parameters across a large range. In our simulations, the upper-leaflet is populated with 11.1% concentration of GPI-AP and rest by equal amounts of $lo$ and $ld$ components. In the lower leaflet, we have 22.2% occupied by PS and other $lo$ components and the rest by the $ld$ component. The TMABD concentration, when present, is kept fixed at 11.1%. This composition approximates the observed composition in a typical patch of membrane in wild-type CHO cells (5).

In our simulations, the interaction energy scale between $lo$ components sets the scale of energy. In these units, the simulations are done at $T = 1.05 T_c$ (i.e. $T > T_c$) where $T_c$ is the critical phase segregation temperature of the $lo - ld$ mixture (3). We set $J_{33} = 0.5 k_B T$ so that PS does not phase segregate in the lower leaflet. In addition, we keep the mutual attraction of GPI-AP and TMABD's small with $J_{11} = J_{66} = 0.15 k_B T$ to avoid any spontaneous clustering. We emphasize that our choice of relative concentration and temperature is not unique and the results hold for a wide range of relative concentrations and temperatures above $T_c$ (6).

To express the results of our simulation in physical units, we denote every length scale in our simulation in units of $\sigma = 10$ nm, a typical molecular scale. This gives us the spatial scale of contractile stress $\xi = 40$ nm and the lateral size of each bilayer patch $L = 600$ nm. We fix the unit of time in our simulations, using a molecular diffusion coefficient of $D \approx 1~\mu\text{m}^2\text{s}^{-1}$ (7), This corresponds to a Monte-Carlo timestep $\Delta t = 0.025$ ms which is also the time taken by any molecule to travel a distance $\sigma$ at the cell surface. In these units, a typical simulation run of $10^6$ MC steps corresponds to 25 s. The local order parameter $\phi$ is calculated on a two-dimensional grid with spacing $0.03~\mu$m overlaid on the bilayer patch. The simulations are performed with a fixed value of the active radial velocity, implemented as an attempt probability, $\Sigma_0 = 0.8$ (illustrated in Fig. S2b). Majority of our simulations with activity are performed with $A = 30\%$ and $\tau = 1600$ MC steps $= 0.04$ s, unless mentioned otherwise.

## Supplementary Note 3: Details of Experimental Methods

**A. Plasmids, cell culture and labeling.** The following constructs were used (with references): 1) Human GPI-anchored Folate Receptor (FR-GPI) described in detail earlier (8). 2) EGFP-GPI, a chimeric model GPI-AP, where the EGFP was fused to the GPI signal from FR-GPI (9). 3) TMABD/ABD*, a chimeric model transmembrane actin (/ABD* non) binding probe, where the transmembrane form of the folate receptor is fused to a cytosolic actin-binding domain from Ezrin (c-terminal Ezrin ABD) or its mutant non-binding counterpart (2). 4) LactC2-Ez-YFP, a fusion construct of LactC2 domain with the actin-filament binding domain of Ezrin and YFP (10).

Chinese hamster Ovary (CHO) cell lines stably expressing either GPI-anchored GFP (EGFP-GPI, GPI signal from FR-GPI), Folate receptor (FR-GPI) and transmembrane (FR-TMABD/*) constructs were maintained in Ham's F12 media (Hi Media, India) supplemented with 10% fetal bovine serum (FBS, Gibco, USA). PSA3 cell line for phosphatidylserine (PS) perturbation experiments were maintained in Ham's F12 medium in the presence of 10mM ethanolamine and 10% FBS. Cells expressing GFP-tagged membrane proteins were treated with 75 $\mu$g/ml of cycloheximide for 3 hrs at 37°C to clear Golgi associated fluorescence before imaging (11). FR-GPI and FR-TMABD expressing cells were cultured in folic-acid free Ham's F-12 and supplemented with 10% dialyzed FBS to allow efficient labeling with fluorescent analogs of folic acid. N$\alpha$-pteroyl-N$\varepsilon$-BodipyTMR-L-lysine (PLB-TMR) was used to label the cell surface at saturating concentration ($\sim$400 nM) on ice for 1 hr or for 5 mins at 37°C. Transient transfections reported here were carried out $\sim$12$-$16 hours before preparing them for the experiments. The cells were transferred to pre-warmed buffers (150 mM NaCl, 20 mM HEPES, 5 mM KCl, 1 mM $CaCl_2$, 1 mM $MgCl_2$, pH 7.2-7.4 also referred to as Buffer M1; supplement with 2mg/ml glucose, M1-Glc) before placing on a temperature-controlled microscope stage and imaged within 30 min (7, 12).

**B. Treatments and Perturbation.** Cholesterol depletions were carried out by treating the cells with 10 mM M$\beta$CD at 37°C for 30 mins, followed by exogenous labeling of fluorescent probes. Cells were depleted of sphingomyelin by growing cells in culture media containing 40 $\mu$g/ml of Fumonisin B1 (FB1) for 72 hrs at usual culture conditions leading to 40-50% reduction in sphingolipids levels (13). PS depletion (PS-condition) was carried out by growing PSA3 cells in media devoid of ethanolamine (using 10% dialysed FBS) for 36-48 hrs (10). PS levels were restored by growing them in cell culture media with normal serum (PS+ condition). Perturbation of formin activity was carried out by pre-treatment of cells with the formin inhibitor SMIFH2 at 25$\mu$M for 2hr at 37°C. Arp2/3 activity was perturbed by treating cells with CK-666 at 50$\mu$M for 2 hr at 37°C. Unless otherwise mentioned, inhibitors with reversible effects were maintained in the buffer during the imaging session.

**C. Steady-state anisotropy measurements.** Cells singly labeled with PLB-TMR or expressing EGFP-GPI or dually labeled with PLB-TMR and EGFP were imaged using a spinning disk confocal system (Andor Technologies, Belfast, Northern Ireland) custom adapted for fluorescence emission anisotropy (14). The labeled cells were excited with $488/561$ nm laser illumination after which parallel and perpendicular polarized fluorescence emission was resolved by passing the emission tight band-pass emission filters, 500-550 nm (for imaging GFP) and 570-610 nm (for imaging PLB) followed by a nanowire polarizing beam splitter and collected using two EMCCD detectors. Dual color anisotropy imaging was carried out on the same setup by sequential imaging of cells to avoid spectral crosstalk and signal bleedthrough. Image processing, analysis, and quantification were performed using Metamorph 7.0 (Molecular Devices Corporation, CA, USA), MATLAB (Mathworks, USA) and Image J (NIH, USA) as described earlier (14, 15).

**D. Membrane ordering measurements.**

***D.1. Laurdan General Polarisation (GP).*** Laurdan (6-lauryl-2-dimethylamino-napthalene) is a polarity sensitive probe to estimate membrane ordering (or fluidity). Laurdan generalized polarization (GP), a ratiometric measurement based on the fluorescence intensity collected at the two spectral channels allows a quantitative estimation of the membrane order. Laurdan imaging and GP measurements was implemented on the confocal spinning disk anisotropy platform to monitor the protein clustering and membrane order at high spatial resolution. Imaging Laurdan on a confocal spinning disk also offer other unique advantages like comparatively lesser photobleaching and ability to measure GP close to the basal membrane plane and avoid contamination from internal pools of Laurdan. Co-imaging of Laurdan GP and anisotropy necessitated some modifications to existing protocols of Laurdan labeling and imaging (16). Laurdan dye stocks (10 mM) were prepared in DMSO and stored in an airtight and lightproof clean glass vial. CHO cells expressing FR-GPI and TMABD were plated on clean glass coverslip dishes, grown in presence of dialyzed serum (for folate analog binding), and grown for 2 days before imaging was done. The dishes were washed twice gently with M1-glucose buffer before the central well with cells was incubated with labeling mix containing both Laurdan (at 10 $\mu$M) and folate analog PLB (400 nM) at 37°C for 5 mins. This allows quick labeling of cell-surface and prevents the buildup of internal pools of Laurdan associated with labeling Laurdan at 37°C for 30 mins (the usual staining protocol for cells). Post-labeling, the dishes were washed thoroughly to reduce overall background fluorescence in images. Laurdan labeled samples were excited using 405 nm laser line and fluorescence was recorded at the two spectral channels 410-460 nm (Ch1) and 470-530 nm (Ch2). However, co-imaging of Laurdan along with anisotropy imaging requires placing the polarizing beam splitter at the secondary dichroic position. This splits up the emission into two orthogonal polarization components for both the channels. The total intensities for both Ch1 and Ch2 was obtained using similar image analysis procedure used for processing anisotropy images. The background intensities were estimated from M1-Glucose buffer imaged under similar imaging condition as the cells. Additionally, a dilute sample of Laurdan in DMSO ($\sim 1\mu$M) was also imaged at similar conditions to determine the Laurdan g-factor as described earlier (16, 17). The Laurdan GP and g-factor was computed using the following formula

$$GP = \frac{I_{Ch1} - G.I_{Ch2}}{I_{Ch1} + G.I_{Ch2}}$$

where G is the g-factor calculated using,

$$G = \frac{GP_{ref} + GP_{ref}.GP_{mes} - GP_{mes} - 1}{GP_{mes} + GP_{ref}.GP_{mes} - GP_{ref} - 1}$$

Here $GP_{mes}$ is the estimated GP value of the Laurdan in DMSO (the g-factor sample). $GP_{ref}$ is a reference value for Laurdan GP which is conventionally fixed at 0.207 (16). The empirically estimated g-factor of 0.5 was uniformly applied across all experiments to calculate Laurdan GP and obtain spatial Laurdan GP maps of membrane order. Laurdan imaging was carried out at room temperature as at 37°C we saw a rapid equilibration of Laurdan to internal membrane pools and its loss from cell surface (Fig. S5 A, B). The internalization of Laurdan (Fig. S5 A,C and higher temperature (37°C compared to 22°C; Fig. S5 B,D) both led to lowering of the Laurdan GP values (and decrease in dynamic range) as reported earlier (18, 19) and verified independently by us (Fig. S5 A-D).

***D.2. NR12S Blue/Red Ratiometric Imaging.*** The cells (untreated or post-perturbation) were labelled with 1 $\mu$M NR12S (20, 21) in M1-Glc for 5 mins at 37°C, washed gently and imaged in M1-Glc. Cells were imaged on a spinning disk confocal microscope with 100X objective (representative images in Fig. S6 A), illuminated with 561 nm excitation laser and emission output was collected using bandpass filters collecting light in the two spectral windows - 565-585 nm emission filter (blue emission) and 593-643 nm (red emission). The background intensities were estimated from M1-Glc buffer imaged under similar imaging condition as the cells and images were background corrected before computing and quantification of Blue/Red ratiometric images.

**E. Imaging and data analysis.**

***E.1. Quantification of domain abundance and size.*** We have established a systematic and consistent approach in quantifying the features of mesoscale domains and extended the same to study mesoscale organisation of the different cell surface probes (in control and upon perturbations) reported here. Multiple ($\geq$ 20) 6-10 $\mu m^2$ patches were selected from the anisotropy maps of probes (representative image in Fig. S3A) from multiple cells ($\sim 10 - 15$) pooled across independent replicates were pooled to generate a typical pixel anisotropy distribution for each probe (or any given condition). Anisotropy thresholds (typically 1 standard deviation less than peak of the distribution; black line in Fig. S3A for FR-GPI) was set to binarize the maps to show only those pixels with anisotropies below this threshold (Fig. S3A). These pixels are highly enriched in nanoclusters and a criterion of contiguous set of pixels ($\geq$ 5 pixels, 1 pixel: 100 nm or 0.01 $\mu m^2$) was used to define individual mesoscopic domains. The same cut-off that was set for control condition was used for perturbation to compare both. The domains maps were quantified using 'Analyze particles' routines of ImageJ to extract the relative abundance (or domain area fraction) and average domain area (in pixel, area: 0.01 $\mu m^2$) for each analysed patch of the membrane. The data was plotted as notched box-plot and statistically compared using appropriate parametric or non-parametric statistical tests.

***E.2. Analysis of domain size distribution.*** We binarize the thresholded segregation order parameter matrix (simulation) and density profile matrix (experiment). The thresholds selected are shown in Supplemental Table 1. We use MATLAB function bwlabel to detect all connected objects (domain) and count the number of pixels in a given object which is denoted as the domain area $s$. For the experimental data we have considered all objects with $s \geq 3$. This procedure is repeated for many simulated realizations of the segregation order parameter matrix or density profile matrix to generate the probability density function of domain area $P(s)$.

Following analysis described in (3), we fit the domain area distribution to

$$P(s) \sim s^{-\theta} \exp\left[-s/s_0\right],$$

with power law exponent $\theta$ that is exponentially cutoff at a scale $s_0$. Similar domain distributions have been used in other contexts of activity (22, 23). We fit the simulation/experimental $P(s)$ to the above form using a custom MATLAB program to implement a Monte-Carlo method to minimize the residual sum of squares (24).

***E.3. Quantitative relation between clustering (of two probes) and with lipid order.*** We used the following approach to analyze the mesoscale sorting behavior of probes. The two color spatial anisotropy maps of several ($\geq$ 20) 6-10 $\mu$m square membrane patches allowed us to manually identify and compute the anisotropy values of a very small patch of membrane (typically of similar size as the domains of the GPI-AP or TMABD probes). For most cases, we have integrated pixel anisotropies of $4 \times 4$ pixel ROI boxes (pixel size: 100 nm) to calculate the overall average anisotropy of the cluster-enriched and cluster-deficient hotspots on the cells. Anisotropy was computed from ROIs marking the cluster-enriched (hence low anisotropy) and cluster-sparse (high anisotropy) regions from the anisotropy maps obtained for one protein (say FR-GPI in the case of Fig.S4 A-C). The same sets of regions were then transferred to the complementary EGFP-GPI spatial anisotropy map from the same cell, and the anisotropy values of the domains were calculated. Several such ROIs (at least 500) were quantified for both proteins from multiple cells (n=15-20 cells across independent replicates) and the data was pooled to build a scatter plot to show the trend in the data and compute the Pearson's correlation coefficient (R=0.55, p<0.01 between FR- and EGFP-GPI; Fig. S3C) for entire dataset. Next we check the correlation coefficient obtained for each individual probe across two sequential frames of acquisition. A high correlation coefficient (see GFP-GPI T1:T2 and FR-GPI T1:T2 in Fig. S4 D,E) indicates there is significant preservation of domains between two frames and a low correlation values observed between two proteins is not due to the dissolution of the mesoscopic domains over the timescale of imaging ($\sim$ 2 sec). We translated the same approach to analyse the correlation between the values of anisotropy and laurdan GP of domains. First, cluster rich (and poor) domains were identified from the anisotropy maps of the membrane probe (say FR-GPI or TMABD) and the same pool of ROI were transferred to the corresponding Laurdan GP map of the same membrane patch and GP values were extracted (Fig. 4C, S5F). The data is displayed as a scatter plot of GP v/s anisotropy to qualitatively observe the trend and quantify the Pearson's Correlation coefficient (with statistical significance).

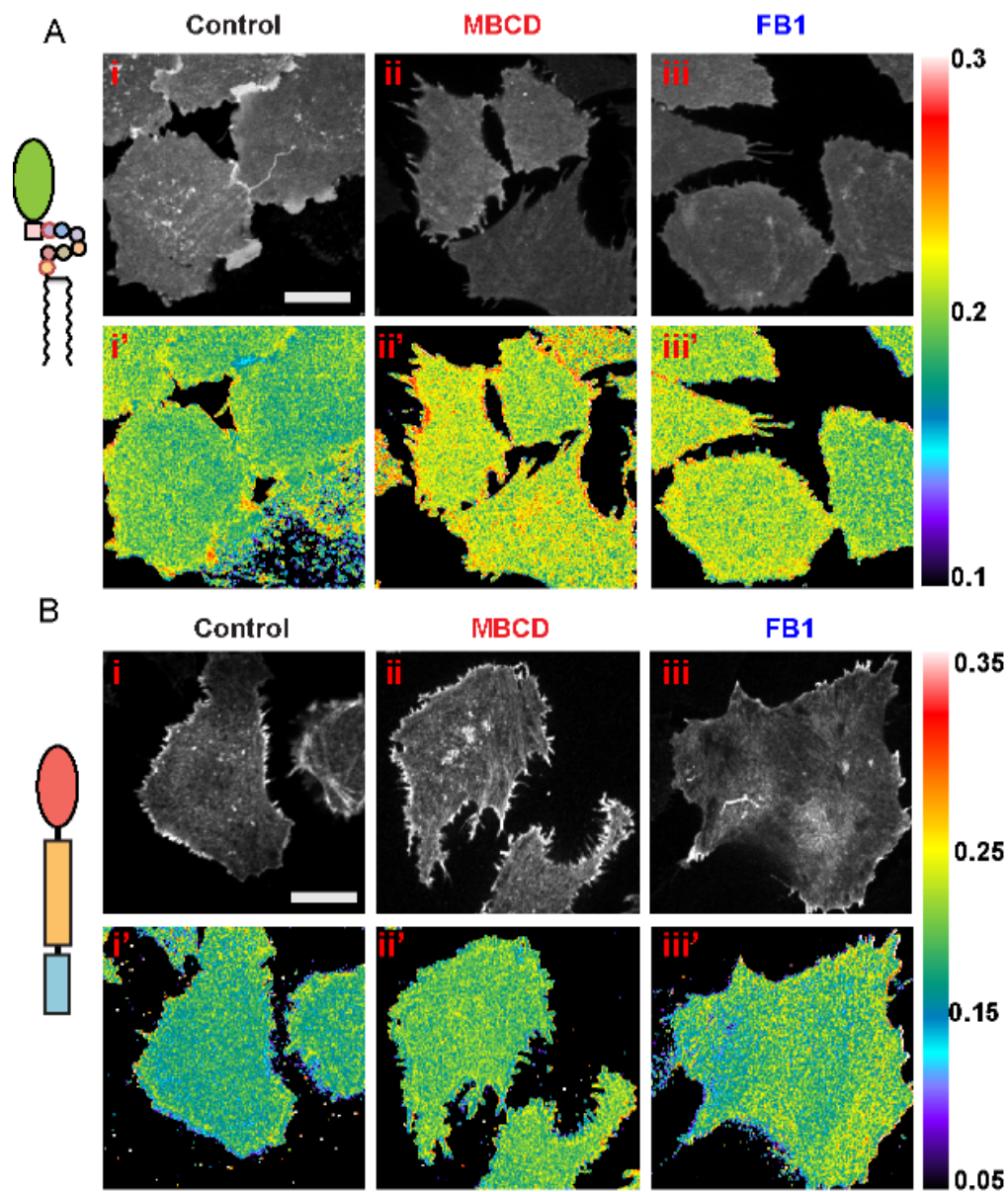

**Fig. S1. Lipid perturbations abrogate clustering of GPI-AP but not TMABD (Related to Figure 1).** (A, B) Representative confocal spinning disk images shows the intensity and anisotropy maps of CHO cells expressing GFP-GPI (A) or FR-TMABD labelled with PLB (B) in untreated (Control), depleted of cholesterol (MBCD) or sphingolipids (Fumonisin B1; FB1). Quantifications of these experiments are presented in Figure 1 F,G. LUT bar indicates the anisotropy scale. Scale bar: 10 $\mu$m.

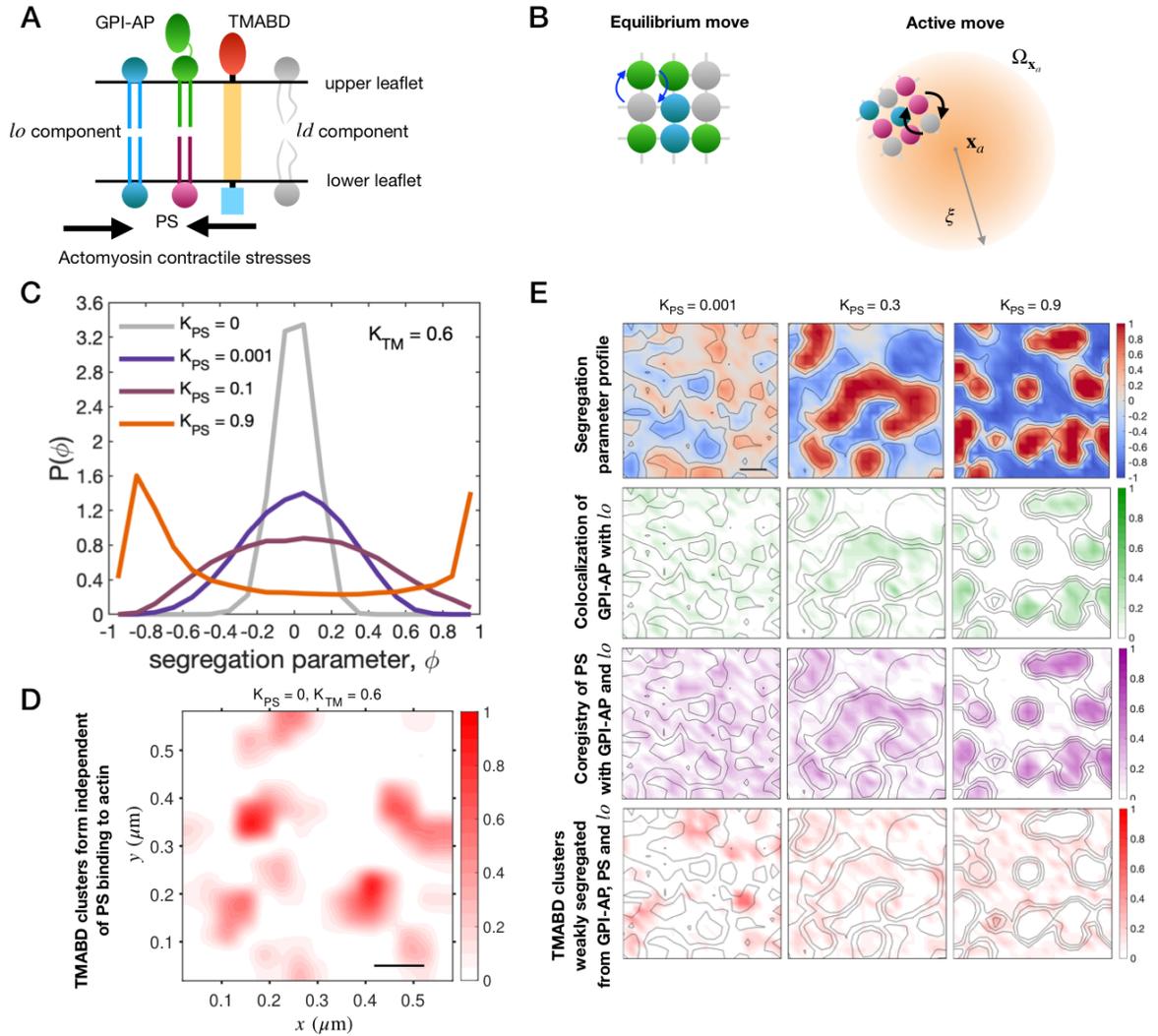

**Fig. S2. Kinetic Monte Carlo simulation model showing active segregation in a multicomponent, asymmetric bilayer.** (a) Schematic of the multicomponent asymmetric bilayer showing various components. (b) Kinetic Monte-Carlo simulation moves: Left – equilibrium moves governed by Kawasaki exchange dynamics (Eq. 4). Right – Active moves on PS and TMABD components when they are within the contractile regions $\Omega_{\mathbf{x}_a}$ and bound to it with a rate $K_{PS}$ and $K_{TM}$, respectively. The active moves result in preferentially advecting the particles towards the centre $\mathbf{x}_a$ with a radial velocity proportional to the contractile stress $\Sigma_0$ (Eq. 3). These moves violate detailed balance. Details of implementation in (3). (c) Probability density functions of $lo$-$ld$ segregation parameter $\phi$ with increasing $K_{PS}$ when TMABD is included in the simulation. We take $K_{TM} = 0.6$. (d) Spatial profile of number density of TMABD calculated following the same protocol as in Fig. 2b or 2e, when PS-actin interaction is switched off ($K_{PS} = 0$). Scale bar 0.1 $\mu$m. (e) Time averaged profiles in the presence of TMABD, of $\phi$ (top row), number densities of GPI-AP (second row), PS (third row) and TMABD (bottom row), at three different values of actin binding affinity of PS.

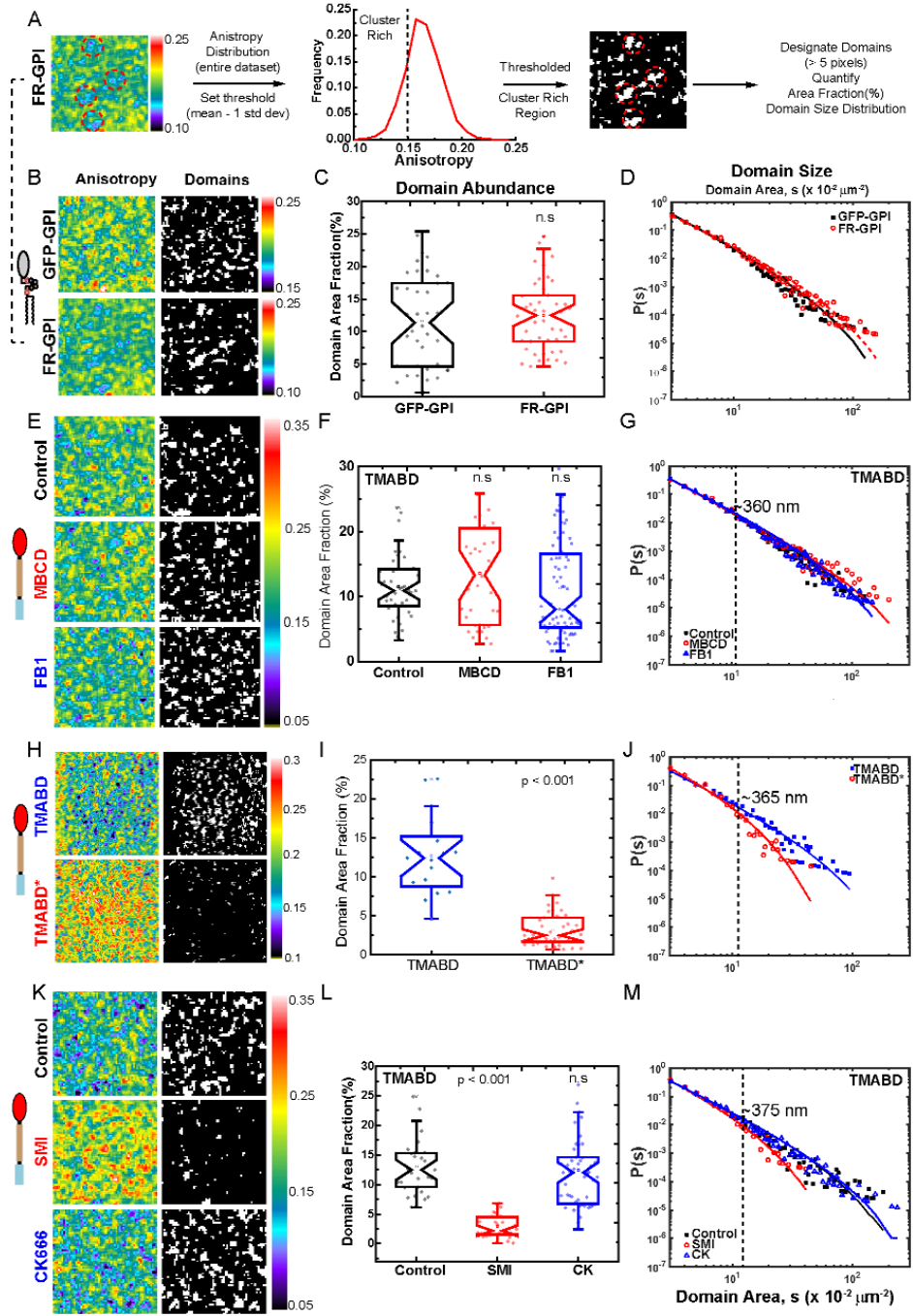

**Fig. S3. Measuring spatial features of mesoscale domains from high-resolution anisotropy maps and the effect of perturbations (related to Figure 3).** A) Analysis approach to measure area fraction and size distribution from anisotropy maps. To generate spatial maps of cluster rich regions ROIs sized $6\ \mu m \times 6\ \mu m$ (left panel) taken from high resolution PLB-labeled FR-GPI anisotropy image, the pixel anisotropy distribution for the entire dataset (middle panel; from at least 30-40 patches derived from 10-20 cells) was used to set an anisotropy cut-off (mean - standard deviation; dashed line in plot). Anisotropy maps were thresholded to obtain binary domain maps (right panel) consisted only of pixels enriched in clusters. We next quantify the abundance (area fraction) and size distribution of individual domains at least 5 pixels in size (red circles on binary maps; > 5 pixel area; pixel : 0.01 um2). (B-D) Representative anisotropy and domain maps of GPI-AP proteins (GFP-GPI or PLB-labelled FR-GPI) from $6\mu m \times 6\ \mu m$ boxes (B) show very similar characteristics of mesoscopic domain abundance (notched box plots in C; note GFP-GPI data used for comparison is the same as depicted in Figure 3A) and distribution of domain area (D); solid lines represent the corresponding fits to a model distribution, $As^{-\theta}exp[-s/s_0]$. (E-G) Representative anisotropy and domains maps (E; $6\ \mu m \times 6\ \mu m$ boxes) of PLB-labelked FR-TMABD in untreated (Control) or depleted of cholesterol (MBCD) or sphingomyelin (FB1) show that both domain abundance (notched box plots in F) and distribution of domain area (G) are insensitive to these perturbations. Fits to domain distribution yield a characteristic domain area of $0.1\ \mu m^2$ which translates to a domain diameter of 360 nm for TMABD (Control, dashed line on plot G). (H-J) Representative anisotropy and domain maps (H; $10\ \mu m \times 10\ \mu m$ boxes) of PLB-labeled FR-TMABD and actin binding mutant version, PLB-TMABD* show that loss of actin interaction capacity results in a significant decrease in domain area fraction (notched box plots in I, two-sample t-test $p < 0.001$) and a marked shift towards smaller domain sizes (J; dashed line on J shows characteristic domain diameter of $\sim 365$ nm). (K-M) Representative anisotropy and domain maps (K; $6\ \mu m \times 6\ \mu m$ boxes) for PLB-labeled FR-TMABD in untreated (Control) or cells treated with inhibitors of formins (SMI; red) and Arp2/3 (CK666; blue) show that formin inhibition but not Arp2/3 led to significant drop in TMABD domain abundance (notched box plots in L; $p < 0.001$ by one-way ANOVA with Tukey means comparison) and a shift towards smaller domain sizes (M; dashed line on M indicates domain diameter of $\sim 375$ nm). The dataset presented here is pooled from multiple independent replicates (2-3) comprising of at-least n = 20-50 anisotropy patches collected from $10 - 20$ cells for each condition in each replicate

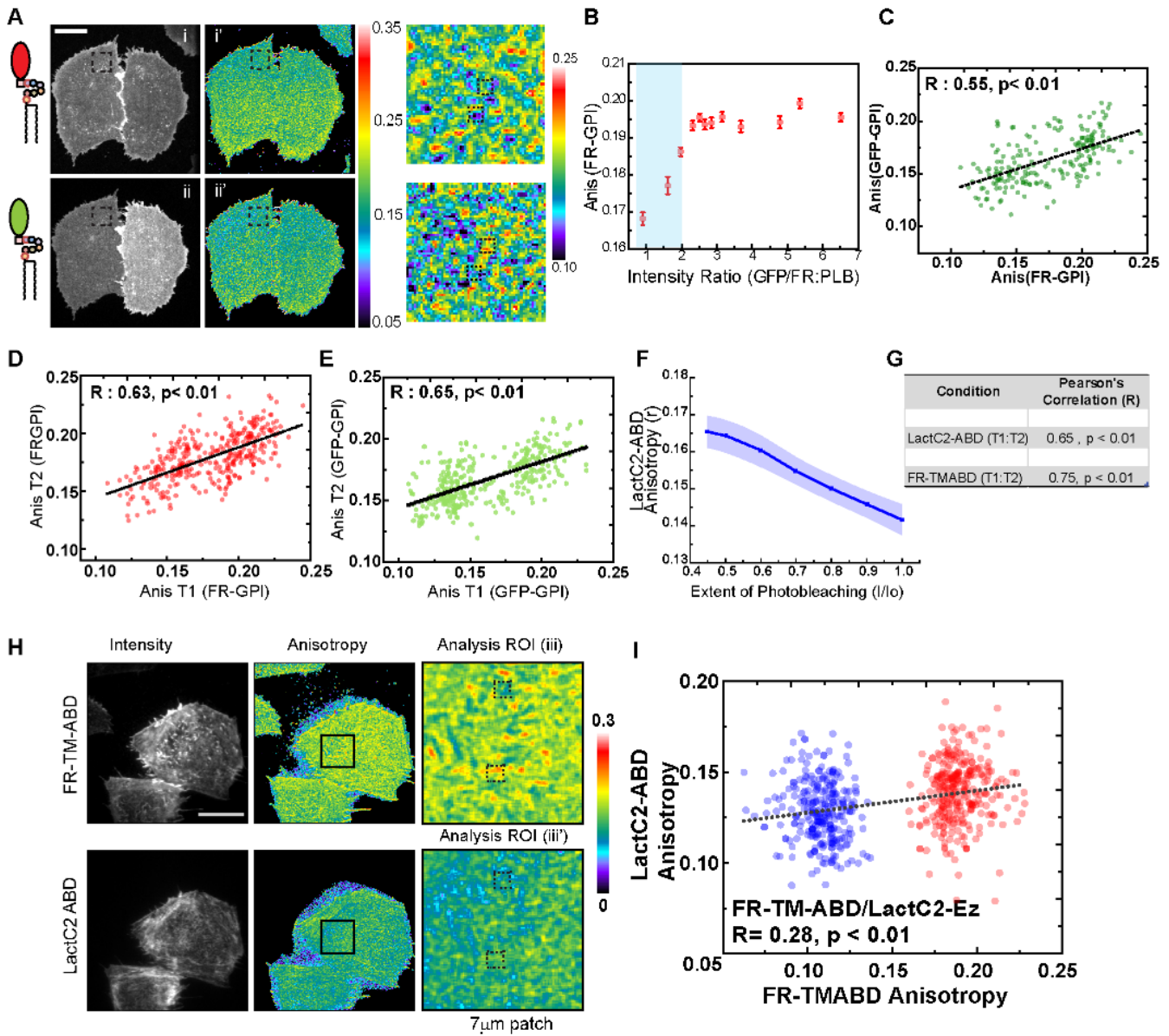

**Fig. S4. Correlation of properties of mesoscopic regions (Related to Figure 4)**. A-E) CHO cells stably expressing FR-GPI (and labeled with PLB; A, i) and transiently expressing GFP-GPI (A, ii) were imaged using a spinning disk confocal microscope to generate representative intensity and anisotropy maps of PLB-labeled FR-GPI (A i') and GFP-GPI (A ii'). The expression of two GPI-APs in the same membrane at comparable levels dilutes the nanoclusters of the individual proteins. To select the region of relative expression levels where the contribution of Homo-FRET dilution is minimal, the anisotropy of PLB-labeled FR-GPI is plotted against GFP/PLB ratio (B). The blue-band (B) outlines the regime where the two probes form co-clusters, and was used to cells co-expressing appropriate levels of FR-GPI and GFP-GPI to probe the colocalisation of cluster rich domains of GFP-GPI and FR-GPI. A $6\mu m \times 6\mu m$ patch (right panel) from the anisotropy maps of FR-GPI (top) and EGFP-GPI (bottom) was outlined from the same cell (dashed box in A i'and ii'; right panel). Mean anisotropy values quantified from corresponding ROIs (4x4 pixels, dashed boxes) on both images were used to generate scatter plots of anisotropy values of PLB-labeled FR-GPI and GFP-GPI (C). Colocalisation of the anisotropy values of these two probes is indicated by a significant positive correlation correlation (Pearson's Correlation coefficient of 0.55, $p < 0.01$). Scatter plot of anisotropy from a similar analysis of two sequential frames of the same GPI-AP probe; PLB-labeled FR-GPI (D) and GFP-GPI (E) shows a high degree of correlation, indicating a preservation of domain maps over the timescale of the sequential multi-colour imaging routine (2-3 sec). F) YFP-LactC2-ABD, a synthetic linker construct that links PS (via the LacC2 domain) to cortical actin (via the ABD from Ezrin) was expressed in CHO cells and subject to photobleaching while collecting images for anisotropy imaging. Photobleaching led to a gradual increase in anisotropy (+/- sd) (data from n = 11 cells), consistent with YFP-LactC2-ABD exhibiting significant homo-FRET as a consequence of its nanoscale clustering. G) Table shows Pearson's correlation coefficient values taken from scatter plot of anisotropy data from a anisotropy correlation analysis (analyzed as in D, E) of two sequential frames of YFP-LactC2-ABD or PLB-labeled FR-TMABD . Note these are also well correlated across two consecutive frames of imaging the same fluorescent molecule. (H, I) Representative TIRF images of fluorescence intensity and anisotropy (H) of cells expressing FR-TMABD (labeled with PLB; upper panel in H) and transiently transfected with YFP LactC2-ABD (lower panel in H). Identical $7\mu m \times 7\mu m$ sized patches (black boxes in H, middle panel) chosen from the anisotropy images of FR-TM-ABD and LactC2-Ez and shown as Analysis ROI (iii, iii'). Scatter plots (I) exhibit significant positive correlation (R = 0.28, $p < 0.01$) between the anisotropy of small mesoscale domain sized ROIs ($400 \times 400 nm$) drawn around cluster-rich and cluster-poor hotspots of FR-TMABD and LactC2-ABD (H; iii, iii', dashed boxes). Scatter plot (I) report on 600 ROIs collected from cells (n=21) pooled from two independent replicates. Scale bar for images (A, H) : $10\mu m$.

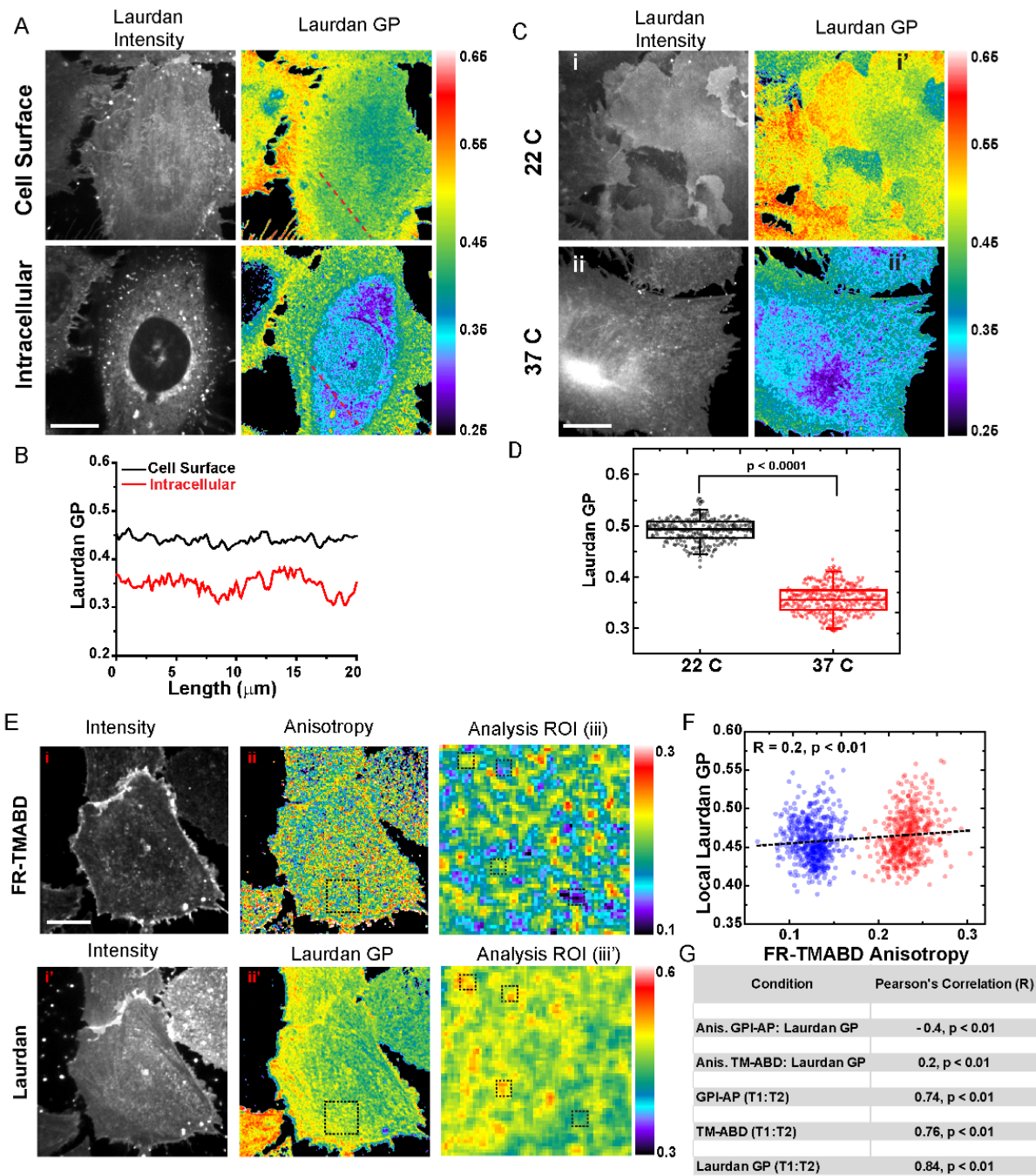

**Fig. S5. Membrane order measurements in living cells using Laurdan (Related to Figure 4).** (A, B) Laurdan intensity and GP image (A) of same cell (also represented in Fig. 5B) at the basal surface (cell surface) and an internal plane (intracellular). Internal membrane pools show distinct Laurdan staining, but register significantly lower GP values, as shown in graph (B) of the data derived from the dashed line in the laurdan GP image. The dynamic ranges of Laurdan GP reported in Figures 5 clearly indicate that the measurements are specific to the cell-surface pool and hence faithfully report on cell membrane lipid ordering. (C, D) Laurdan GP distribution is extremely sensitive to temperature of labelling and measurement. At $22°C$, Laurdan intensity images shows more prominent cell membrane specific labeling (C i) but $37°C$, it is internalized rapidly and enters the cell, as evident here (C ii). Moreover, the Laurdan GP values reduce drastically at higher temperatures (C ii') as quantified in box-plot (D). Therefore the Laurdan GP measurements were conducted at room temperature ($22°C$). E-G) High-resolution confocal fluorescence and anisotropy images of CHO cells expressing FR-TMABD (labeled with PLB, E, top panel) and co-labelled with Laurdan (E, bottom panel, Intensity) to obtain Laurdan GP maps. Identical $6 \mu m x 6 \mu m$ sized boxes (black squares) were chosen across the anisotropy and Laurdan GP map and shown on the right (Analysis ROI; E iii, iii', respectively). Scatter plot (F) of local Laurdan GP and anisotropy computed for small mesoscale domain ($400 \times 400$ nm) from identical ROIs depicted in the Analysis ROI for PLB-labeled FR-TMABD (E iii, dashed boxes) and Laurdan GP (E iii', dashed boxes). Laurdan GP and anisotropy of PLB FR-TMABD show a small but significant positive correlation ( Pearson's R = 0.2, $p < 0.01$). Table (in G) shows Pearson's correlation coefficients (R) obtained from scatter plots of anisotropy values of PLB-labeled FR-GPI-AP or PLB-labeled TMABD against their local Laurdan GP (rows, 1, 2). Note high degree of correlation of clustering of GPI (row 3), TMABD (row 4) and local lipid order (Laurdan GP; row 5) across two sequential frames. Individual scatter plot (F) report on at least 500 ROIs collected from cells (n=20 cells) pooled from two independent replicates. Scale bar for images (A,C & E) : $10 \mu m$.

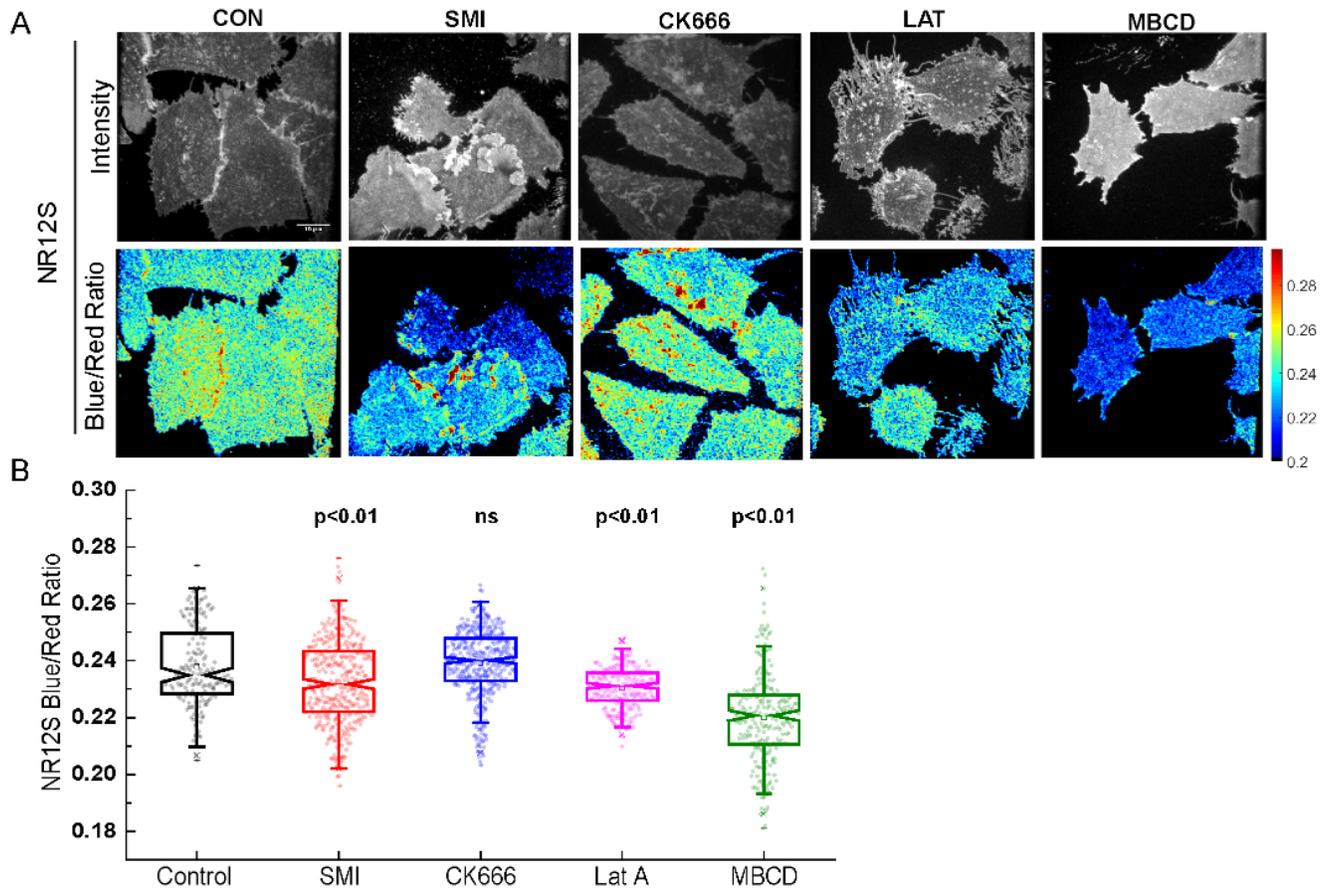

**Fig. S6. Steady-state lipid order of the outer-leaflet of cell surface probed by solvatochromic probe, NR12S (Related to Figure 5)** A) Confocal images showing the intensity and Blue/Red Ratio images of CHO cells labelled with 1 $\mu$M NR12S in untreated (CON/control) condition or upon perturbations of formins (SMI), Arp2/3 (CK), actin polymerization (LAT) and cholesterol (MBCD). Steady-state outer-leaflet membrane order are reflected in the ratio changes observed in these images (quantified in B; scale bar : 10 $\mu$m). B) Box-plots of NR12S Blue/Red ratio for the different perturbation conditions. Perturbation of Formin (SMI) and Latrunculin (LatA) but not Arp2/3 (CK666) lead to loss of outer-leaflet membrane order. The difference between untreated (Control) and cholesterol depleted (MBCD) cells establish the dynamic range captured by NR12S in this assay. At-least 200 ROIs were considered for the analysis from 40-50 cells each condition across 2 independent experiments. Statistical significance of difference between control and each perturbation was determined by one-way ANOVA with Tukey mean comparison.

**Supplemental Table 1: Fit Parameters for Domain Size Distribution**

| Experiment | Membrane Probe | Fitted Values ± errors | | |
|---|---|---|---|---|
| | | A | θ | s0 (x 0.01 μm²) |
| **Tuning f-actin affinity** | | | | |
| F-actin binding | TM-ABD | 3.634 ± 0.020 | 2.105 ± 0.002 | 38.394 ± 0.931 |
| No F-actin binding | TM-ABD* | 5.181 ± 0.067 | 1.917 ± 0.008 | 7.305 ± 0.057 |
| **Lipid Perturbations** | | | | |
| GPI-AP control | GPI | 4.264 ± 0.097 | 2.205 ± 0.003 | 40.412 ± 1.758 |
| GPI-AP Sphingomyelin Depletion | GPI (FB1) | 10.881 ± 0.426 | 2.380 ± 0.041 | 7.260 ± 0.654 |
| GPI-AP Cholesterol Depletion | GPI (MBCD) | 18.615 ± 1.357 | 2.430 ± 0.115 | 4.059 ± 0.552 |
| TM-ABD control | TM-ABD | 4.855 ± 0.024 | 2.287 ± 0.003 | 73.689 ± 3.130 |
| TM-ABD Sphingomyelin Depletion | TM-ABD (FB1) | 3.922 ± 0.020 | 2.106 ± 0.002 | 46.650 ± 2.298 |
| TM-ABD Cholesterol Depletion | TM-ABD (MBCD) | 4.142 ± 0.015 | 2.203 ± 0.001 | 93.551 ± 2.701 |
| **Formin Perturbations** | | | | |
| GPI Control | GPI | 3.263 ± 0.021 | 2.004 ± 0.003 | 46.127 ± 2.950 |
| GPI Formin Inhibition | GPI (SMIFH2) | 8.657 ± 0.574 | 2.236 ± 0.073 | 7.175 ± 0.384 |
| TM-ABD control | TM-ABD | 3.395 ± 0.021 | 2.005 ± 0.002 | 37.611 ± 1.801 |
| TM-ABD Formin Inhibition | TM-ABD (SMIFH2) | 4.557 ± 0.282 | 2.017 ± 0. 012 | 10.752 ± 0.106 |
| **Two GPI Comparison** | | | | |
| GFP-GPI | | 3.263 ± 0.021 | 2.205 ± 0.003 | 46.127 ± 2.950 |
| FR-GPI | | 4.264 ± 0.097 | 2.004 ± 0.003 | 40.412 ± 1.758 |
| | | | | |
| **Simulation** | **Tunable Parameter** | **A** | **θ** | **s0** |
| Tuning Binding Affinity of PS | K_PS | | | |
| | 0.9 | 0.561 ± 0.001 | 1.602 ± 0.003 | 17.919 ± 0.049 |
| | 0.3 | 0.700 ± 0.001 | 1.795 ± 0.003 | 11.876 ± 0.075 |
| | 0.1 | 0.849 ± 0.001 | 1.995 ± 0.003 | 6.939 ± 0.033 |
| | 0.001 | 1.126 ± 0.007 | 1.939 ± 0.010 | 2.905 ± 0.050 |
| Tuning Area Fraction of Aster | A* | | | |
| | 0.07 | 0.799 ± 0.001 | 1.994 ± 0.003 | 9.896 ± 0.057 |
| | 0.154 | 0.650 ± 0.001 | 1.599 ± 0.001 | 13.521 ± 0.437 |
| | 0.238 | 0.568 ± 0.001 | 1.600 ± 0.001 | 18.289 ± 0.558 |
| | 0.349 | 0.550 ± 0.001 | 1.593 ± 0.007 | 19.864 ± 0.042 |
| | | | | |
| Note: All distributions are generated by thresholding binary images for a domain size >= 3 pixels. | | | | |